\documentclass[aip,reprint]{revtex4-1}

\draft 

\usepackage{graphicx}	
\usepackage{amsmath}	
\usepackage{amssymb}
\usepackage{bm}
\usepackage{verbatim}
\usepackage{xcolor}
\usepackage{hyperref}
\usepackage{mathtools}
\usepackage{amsfonts}
\usepackage{graphics}
\usepackage{listings}
\usepackage[normalem]{ulem}      
\usepackage{color,soul}
\newcommand\util{\tilde{u}_{\textbf{k}, \omega}}
\newcommand\ztil{\tilde{\zeta}_{\textbf{k}, \omega}}
\newcommand\pderkx{\frac{\partial}{\partial k_x}}

\newcommand\pdery{\frac{\partial}{\partial y}}
\newcommand\pderz{\frac{\partial}{\partial z}}

\newcommand\Btil{\tilde{B}_{x;\textbf{k}, \omega}}
\newcommand\zBtil{\tilde{\zeta}_{B;\textbf{k}, \omega}}
\newcommand\BB{\mathcal{B}}

\begin{document}


\title{The {competition} between the hydrodynamic instability from noise and magnetorotational 
instability in the Keplerian 
disks} 



\author{Subham Ghosh}
\email[]{subham@iisc.ac.in}
\affiliation{Department of Physics, Indian Institute of Science,
Bangalore 560012, India}

\author{Banibrata Mukhopadhyay}
\email[]{bm@iisc.ac.in}
\affiliation{Department of Physics, Indian Institute of Science,
Bangalore 560012, India}



\setstcolor{red}

\begin{abstract}
\noindent
We venture for the comparison between growth rates for magnetorotational instability (MRI) and hydrodynamics instability 
in the presence of an extra force in the local Keplerian accretion flow. The underlying model is described by the 
Orr-Sommerfeld and Squire equations in the presence of rotation, magnetic field and an extra force, plausibly noise with 
a 
nonzero mean. We obtain MRI using Wentzel-Kramers-Brillouin (WKB) approximation without extra force for purely vertical 
magnetic field and vertical 
wavevector of the perturbations. Expectedly, MRI is active within a range of magnetic field, which changes depending on 
the perturbation wavevector magnitude. Next, to check the effect of noise on the growth rates, a quartic dispersion 
relation has been obtained. Among those four solutions for growth rate, the one that reduces to MRI growth rate at the 
limit of vanishing mean of noise in the MRI active region of the magnetic field, is mostly dominated by MRI. However, 
in MRI inactive region, in the presence of noise the solution turns out to be unstable, which are almost independent of 
the magnetic field. Another growth rate, which is almost complementary to the previous one, leads to stability at the 
limit of vanishing noise. The remaining two growth rates, which correspond to the hydrodynamical growth rates at the 
limit 
of the vanishing magnetic field, are completely different from the MRI growth rate. More interestingly, the latter growth 
rates are larger than that of the MRI. If we consider viscosity, the growth rates decrease depending on the Reynolds 
number.
\end{abstract}


\maketitle 

\section{Introduction}
\label{sec:intro}

The accretion disk is one of the exotic objects in Astrophysics. As the name suggests, they form in a disk-like shape due 
to the accretion of matter {with angular momentum} around a center by unleashing enormous gravitational energy 
through radiation. They are ubiquitous. From around the center of the galaxies to the formation of planets, from the 
binary system to the ring of Saturn, they prevail at various length scales. The physics related to the accretion disks is 
also exotic. There have been a lot of studies regarding the phenomena related to the accretion disks, and it still 
continues. However, still {there are} some open questions in the field of accretion disks. One of those is how, 
in the first place, the accretion disk forms, i.e., how the matter falls in, losing the angular momentum and eventually 
reaches the central object{, particularly where the molecular viscosity is negligible}. We attempt to address this 
question in this work.

To match with the observational evidences \cite{Frank_2002}, the accretion flow has to be turbulent \cite{Shakura_1972}. 
However, the accretion flow in the Keplerian accretion disk, where the centrifugal force almost balances the gravity, is 
Rayleigh (centrifugally) stable, i.e., the infinitesimal perturbation eventually dies down. The route to turbulence, 
therefore, is essential to find. In 1991, \citealt{Balbus_1991} proposed the idea of an ideal magnetohydrodynamical (MHD)
instability following \citealt{Velikhov_1959} and \citealt{Chandrasekhar_1960}: magnetorotational instability (MRI). The 
interplay between the weak magnetic field and the rotation of the fully ionized plasma gives rise to the MRI. It is a 
widespread mechanism in {leading to turbulence and} explaining the transport of angular momentum in accretion disks. 
{This is} because it grows within the dynamical timescale with very minimum requirements: weak magnetic field so that 
the corresponding magnetic pressure is smaller than the hydrodynamical pressure, and a radially decreasing angular 
velocity in contrast {to} the radially decreasing specific angular momentum as the Rayleigh stability criterion 
\cite{Balbus_1998}.  Despite of its popularity in the accretion disk community and tremendous success in explaining 
various accretion phenomena over the year{s}, the MRI is not out of caveats. When the temperature is low, the fluid 
and the magnetic field do not get coupled well due to inefficient ionization. Consequently, the nonideal MHD effects come 
into the picture. Due to the stabilizing effects of {O}hmic resistivity on the linear instability, there are many 
regions in protoplanetary disks where MRI gets suppressed \cite{Gammie_1996, Fleming_2003}. The additional inclusion of 
ambipolar diffusion makes this scenario worse \cite{Turner_2014}. 

In order to explain the angular momentum 
transport, magnetised disk winds have been considered {as} an alternative to MRI turbulence\cite{Bai_2013_ApJ_a, 
Bai_2013_ApJ_b}. If the toroidal component of the magnetic field is beyond the critical value, MRI gets suppressed 
locally\cite{Pessah_2005} and globally \cite{Das_2018}. Instead, there arise different MHD instabilities. 
{Moreover,} due to the nonnormality of the underlying system, it has been argued that the magnetic transient 
growth\cite{Nath_2015} brings nonlinearly and hence plausible turbulence faster than MRI for the flows with the Reynolds 
number ($Re$) beyond $10^9$. Since $Re$ in accretion disk\cite{Mukhopadhyay_2013} is much larger than this value, MRI is 
questionable in systems {with large $Re$}. It, therefore, would be an enthralling venture to look for the instability 
mechanisms of 
hydrodynamical origins.

A lot of efforts have been put forward to understand the angular momentum transport in accretion disks in the laboratory, 
i.e., through {the table top} experiments and simulations. To do so, the accretion disk-like environment, i.e., the 
Keplerian angular velocity, has been created in the experiments. In the experimental context, the Taylor-Couette flow 
\cite{Chandrasekhar_1961} has been considered. Its crude description is the following. It has two concentric cylinders 
and fluid in between. By moving the cylinders, the fluid is kept in motion. The flow parameters, i.e., the velocity, 
angular velocity, $Re$, etc., are controlled by the speed of both the cylinders\cite{Landau_1959}. Although 
the exploration of the transport of angular momentum in accretion disk through laboratory started by 
\citealt{Richard_1999} in 1999, based on the {key} experimental results, there are mainly two groups in the 
community: one (Group 1) who obtained {hydrodynamic} turbulence \cite{van_Gils_2011, Paoletti_2011, Paoletti_2012} 
for the Keplerian angular velocity profile and the other one (Group 2) who did not obtain {any hydrodynamic} 
turbulence\cite{Ji_2006} for the same. All the experiments reached the equivalently same $Re$ to draw their conflicting 
conclusions. Although they obtained the same $Re$, the two groups have different experimental devices. Group 1 did the 
experiment with {a} tall and narrow device, while Group 2 used {a} small and wide apparatus. These kinds of 
geometrical differences in the apparatus affect the critical $Re$ of the corresponding flows remarkably 
\cite{Richard_1999, Fromang_2019}. For the same reason, turbulence is hard to reach in the case of wide apparatus. It 
could be plausible for Group 2 that they operated the apparatus at the verge of the critical $Re$ required for the 
transition to turbulence or below the critical $Re$ so that they did not reach the turbulent regime. Although, in 
reality, there is no rigid axial boundary in accretion disks, due to the finite size of the apparatus, each group 
differently implemented the axial boundaries. To check the effect of these axial boundaries on the flow, a direct 
numerical simulation \cite{Avila_2012}, with very small $Re$ compared to the experiments due to the computational 
constraint, was performed {in the equivalent set up of} the experiments done by both the groups. The simulations of 
both the experiments show that the flows become turbulent mainly due to the presence of the boundaries. Hence, the 
effects of boundaries, i.e., the finite size of the apparatus have important roles in bringing turbulence. From the 
theoretical, numerical, and experimental points of view, the current status of the field is still far from {settling} 
down. 

Keeping in mind the nonmagnetic Keplerian flow, an alternative approach of linear transient growth 
\cite{Tevzadze_2003, Chagelishvili_2003, Yecko_2004, man_2005, amn_2005} {has been also} proposed. Due to the 
{non-selfadjoint property} of the underlying differential operator, the total energy \cite{Schmid_2002} of the linear 
perturbations in the shear flow becomes significantly large for a short time before {their} eventual decay. Although 
its tremendous success is in explaining the subcritical transition to turbulence in the laboratory shear flows, 
authors
\cite{man_2005, Lesur_2005, Shi_2017} {questioned} for the corresponding sustained turbulence in the Keplerian flow. 
However, several orders of magnitude discrepancies exist between the current computational resolution and the $Re$ of the 
accretion disk. Nevertheless, we encounter the problem from the hydrodynamical point of view but in the presence of an 
extra force. Considering this extra force to be the white noise with a zero mean, the multi-mode analysis of the 
Keplerian flow has been studied\cite{Ioannou_2001, Raz_2020}. We instead follow the idea and methodology put forwarded by 
Mukhopadhyay and Chattopadhyay \cite{Mukhopadhyay_2013}.  Recently, authors\cite{Nath_2016, Ghosh_2020, Ghosh_2021_a, 
Ghosh_2021_b} argued that if the white noise has a nonzero mean, the Keplerian flow effectively becomes linearly 
unstable. 
{The origin of the extra force in the context of accretion flow, 
as merely mentioned by Ioannou and Kakouris \cite{Ioannou_2001}, could be the neglected nonlinear terms during the linear 
analysis, supernova explosions, etc. However, the other plausible models for its origin, as 
proposed explicitly and rigorously by Ghosh and Mukhopadhyay\cite{Ghosh_2020} very recently, could be the dust-grain 
interaction in protoplanetary disks, feedback from the outflow/jet on the accretion disks, etc.} 
Nevertheless, the comparison 
between the growth rates of hydrodynamical 
instability in the presence of {an extra force} and that of the MRI has never been studied. {Although 
there could be different models for the extra force\cite{Ioannou_2001, Ghosh_2020}, we however, restrict 
ourselves to the white noise with nonzero mean as the extra force in this work.} This study has its own significance 
because the presence of noise is not limited \cite{Nath_2016, Ghosh_2020} to the nonmagnetic Keplerian flow. The presence 
of noise, in fact, is quite ubiquitous in {any} Keplerian flow as argued by Nath and Mukhopadhyay\cite{Nath_2016} and 
later Ghosh and Mukhopadhyay\cite{Ghosh_2020}. Hence, it is very important to compare these two growth rates to check 
whether they come in unison, or one opposes the other. In other words, what the growth rate corresponding to the 
hydrodynamical instability is in the favorable parameter space for MRI. If the growth rate of the hydrodynamical 
instability is comparable to that of the MRI, we can say that irrespective of the magnetic field, the accretion flow is 
linearly unstable. Hence, nonlinearity and turbulence eventually could be the inevitable fate of the magnetic or 
nonmagnetic Keplerian flow. 

The decoration of the paper is the following. In \S\ref{sec:Formalism}, {we} first formulate the problem in the local 
shearing box situated at a local patch of the accretion disk. In the local box, the governing equations to describe the 
linearly perturbed background flow are the magnetic Orr-Sommerfeld and Squire equations in the presence of an extra 
force and the Coriolis force. The corresponding background magnetic and velocity fields are also described here. With 
these equations, in the absence of the extra force, we obtain a dispersion relation neglecting the hydrodynamic and 
magnetic viscosities to obtain the criteria for centrifugal {instability} (CI) and {MRI} in 
\S\ref{sec:CI_MRI}. The effect of the 
{force} is described in \S\ref{sec:effect_of_noise}, where the dispersion relation in the presence of noise but in 
the absence of any kinds of viscosity is obtained. {Note that if the external force is not stochastic rather 
deterministic, similar scenario can also be implemented.} The solutions {for} that dispersion 
relation have been studied extensively without magnetic field in \S\ref{sec:without_magenetic_field} and with magnetic 
field in \S\ref{sec:With_magnetic_field}. In \S\ref{sec:the_effect_of_Re}, we include the effect of the hydrodynamic 
viscosity on the growth rates. We conclude in \S\ref{sec:Conclusion} that depending on the extra force, the hydrodynamic 
growth rate in the favorable parameter zone for MRI becomes greater than that of the MRI. For other parameters, which 
lead to the suppression of MRI, it is the hydrodynamical instability that makes the flow unstable and plausibly further 
turbulent. Hence, the presence of noise in the flow is enough to make the underlying flow unstable.

\section{Formalism}
\label{sec:Formalism}
We refer to the local patch of the Keplerian accretion disk, where we do the whole analysis. For the detail{ed} 
description of the local formulation, please see \cite{man_2005, Bhatia_2016, Mukhopadhyay_2011NJPh}. The governing 
equations of the perturbed flow are magnetic Navier-Stokes equation in the presence of rotation and noise, the 
magnetic induction equation with the constraints of incompressibility \cite{Nath_2015, amn_2005} due to the local nature 
of the flow, and the no-magnetic monopole. We, however, recast the governing equations into magnetic Orr-Sommerfeld and 
Squire equations ({see} Appendix of {\citealt{Mukhopadhyay_2013}} for the 
detail{ed} 
derivation). 
{The background and perturbed velocities are $\textbf U_0 = 
(0,-x,0)${\cite{Ghosh_2021_b, man_2005, Balbus_1998}} and $(u,v,w)$, respectively. For the present
purpose, the background and 
perturbed magnetic fields are $\textbf{B}_0 = (B_{0x}, B_{0y}, B_{0z})$ and $(B_x, B_y, B_z)$, respectively. 
Hence, taking into 
account the above mentioned background and perturbed quantities, the magnetized Orr-Sommerfeld and Squire equations 
become}
\begin{subequations}
 \begin{eqnarray}
 \begin{split}
   \left(\frac{\partial}{\partial t}-x\frac{\partial}{\partial y}\right)\nabla^2u 
   + \frac{2}{q}\frac{\partial \zeta}{\partial z} - \frac{1}{4\pi}\left(\textbf{B}_0\cdot \bm{\nabla}\right) \nabla^2B_x
   \\= \frac{1}{Re}\nabla^4u +\eta_1,
   \label{eq:gen_velo_pertb}
   \end{split}
   \end{eqnarray}
   \begin{eqnarray}
 \begin{split}
   \left(\frac{\partial}{\partial t}-x\pdery\right)\zeta +\left(1 - \frac{2}{q}\right) 
\frac{\partial u}{\partial z} - \frac{1}{4\pi}\left(\textbf{B}_0\cdot \bm{\nabla}\right) \zeta_B \\= \frac{1}{Re}\nabla^2 
\zeta 
+\eta_2,
\label{eq:gen_vorti_pertb}
 \end{split}
\end{eqnarray}
 \begin{eqnarray}
 \begin{split}
 \left(\frac{\partial}{\partial t}-x\pdery\right)B_x - \left(\textbf{B}_0\cdot \bm{\nabla}\right)u = \frac{1}{Rm}\nabla^2 
B_x,
 \label{eq:gen_mag_velo_pertb}
 \end{split}
\end{eqnarray}
\begin{eqnarray}
 \begin{split}
 \left(\frac{\partial}{\partial t}-x\pdery\right)\zeta_B - \left(\textbf{B}_0\cdot \bm{\nabla}\right)\zeta - \pderz B_x
 \\= \frac{1}{Rm}\nabla^2\zeta_B, 
 \label{eq:gen_mag_vorti_pertb}
 \end{split}
\end{eqnarray}
\end{subequations}
where $\eta_{1,2}$ are the extra force; $\zeta = \partial w/\partial y- \partial v/\partial z$ and $\zeta_B= \partial 
B_z/\partial y- \partial B_y/\partial z$ are the $x$-component of vorticity and magnetic vorticity, respectively; $Re$ 
and $Rm$ are the Reynolds and magnetic Reynolds numbers, respectively; $q$ is the rotation parameter describing the 
radial dependence ($r$) of the angular frequency ($\Omega$) of fluid element around the central object, given
by $\Omega \propto r^{-q}$.

We have to write down all the equations (\ref{eq:gen_velo_pertb}), (\ref{eq:gen_vorti_pertb}), 
(\ref{eq:gen_mag_velo_pertb}) and (\ref{eq:gen_mag_vorti_pertb}) in the Fourier space in the due course of calculation. 
For that our convention of the Fourier {transformation} and the inverse Fourier {transformation} are respectively,
\begin{subequations}
\begin{equation}
 A(\textbf{r},t) = \int{\tilde{A}_{\textbf{k},\omega}}e^{i(\textbf{k}\cdot \textbf{r} - \omega t)}d^3k\ d\omega
 \label{eq:Fourier_trans}
\end{equation} 
and
\begin{equation}
 \tilde{A}_{\textbf{k},\omega} = \left(\frac{1}{2\pi}\right)^4\int{A(\textbf{r},t)e^{-i(\textbf{k}\cdot \textbf{r} - 
\omega t)}} d^3x\ dt,
 \label{eq:inv_Fourier_trans}
\end{equation}
\end{subequations}
where $A$ can be any one of $u, \zeta, B_x, \zeta_B, \eta_1\ \rm {and}\ \eta_2;$ $\textbf{k}$ and $\omega$ are the 
wavevector and the corresponding frequency of the perturbation in the Fourier 
space such that in Cartesian coordinates $\textbf{k}{\equiv}(k_x,k_y,k_z)$ and $|\textbf{k}| = k;$ $\textbf{r}$ is 
the position vector and in Cartesian coordinates $\textbf{r} {\equiv} (x,y,z).$

The boundary conditions to solve equations (\ref{eq:gen_velo_pertb}), (\ref{eq:gen_vorti_pertb}), 
(\ref{eq:gen_mag_velo_pertb}) and (\ref{eq:gen_mag_vorti_pertb}) are 
\begin{equation}
 u=\frac{\partial{u}}{\partial{x}} = \zeta = B_x = \frac{\partial{B_x}}{\partial{x}} = \zeta_B = 0, \ {\rm at}\ x = \pm 
1. 
 \label{eq:boundary_condition} 
\end{equation}

In the Fourier space, the above equations {reduce to} 
\begin{subequations}
 \begin{eqnarray}
 \begin{split}
- k_yk^2 \frac{\partial  \tilde{u}_{\textbf{k},\omega}}{\partial k_x} + \left(i\omega k^2 - 2k_xk_y - 
\frac{k^4}{Re}\right) \tilde{u}_{\textbf{k},\omega} \\+\frac{2ik_z}{q} \tilde{\zeta}_{\textbf{k},\omega} 
+ \frac{ik^2}{4 \pi}\left(\textbf{B}_0\cdot\textbf{k}\right)\Btil= \ m_1\delta(\textbf{k})\delta(\omega),
\label{eq:Keplerian_velo_Fourier_space}
\end{split}
\end{eqnarray}
\begin{eqnarray}
 \begin{split}
  k_y \frac{\partial  \tilde{\zeta}_{\textbf{k},\omega}}{\partial k_x} + ik_z \left(1-\frac{2}{q}\right) 
\tilde{u}_{\textbf{k},\omega} + \left(\frac{k^2}{Re} - i\omega \right) \tilde{\zeta}_{\textbf{k},\omega} 
\\- \frac{i}{4 \pi}\left(\textbf{B}_0\cdot\textbf{k}\right)\zBtil  = \ m_2\delta(\textbf{k})\delta(\omega),
 \label{eq:Keplerian_vorti_Fourier_space}
 \end{split}
\end{eqnarray}
\begin{eqnarray}
\begin{split}
 k_y\pderkx\Btil + \left(\frac{k^2}{Rm} -i\omega\right)\Btil \\- i\left(\textbf{B}_0\cdot\textbf{k}\right)\util 
 =0,
 \label{eq:mag_Fourier_space}
 \end{split}
\end{eqnarray}
\begin{eqnarray}
\begin{split}
 k_y\pderkx\zBtil + \left(\frac{k^2}{Rm} -i\omega\right)\zBtil \\- i\left(\textbf{B}_0\cdot\textbf{k}\right)\ztil - 
ik_z\Btil
 =0.
 \label{eq:mag_vorti_Fourier_space}
 \end{split}
\end{eqnarray}
\end{subequations}

{We choose} the solutions of the equations (\ref{eq:gen_velo_pertb}), (\ref{eq:gen_vorti_pertb}), 
(\ref{eq:gen_mag_velo_pertb}) and
(\ref{eq:gen_mag_vorti_pertb}) {be} $\psi = \psi(x)e^{i(\bm{\alpha}\cdot\textbf{r}-\beta t)}$, where $\psi$ can be 
any of 
$u$, $\zeta$, $B_x$ and $\zeta_B$; $\bm{\alpha} = (\alpha_1,\alpha_2,\alpha_3)$ is the wave vector, $\beta$ is the 
frequency. In general, $\beta$ is complex. However, according to our convention, if the imaginary part of $\beta$, i.e., 
$Im(\beta)$ is positive, then the perturbation grows with time and hence makes the flow unstable. In the Fourier space, 
these solutions become 
\begin{eqnarray*}
 \begin{split}
  &\tilde{\psi}_{\textbf{k},\omega} = \left(\frac{1}{2\pi}\right)^4 
\int_{-\infty}^{\infty}{\psi(x)e^{i(\bm{\alpha}\cdot\textbf{r}-\beta t)} 
e^{-i(\textbf{k}\cdot\textbf{r}-\omega t)}}d^3xdt&\\
 =& \frac{1}{2\pi}\delta(\alpha_2-k_y) \delta(\alpha_3-k_z) \delta(\beta - \omega) 
\int_{-\infty}^{\infty}{\psi(x)e^{i(\alpha_1-k_x)x}}dx.&
 \end{split}
\end{eqnarray*}
To obtain the dispersion relation, we integrate the equations (\ref{eq:Keplerian_velo_Fourier_space}), 
(\ref{eq:Keplerian_vorti_Fourier_space}), 
(\ref{eq:mag_Fourier_space}) and (\ref{eq:mag_vorti_Fourier_space})
with respect to $\textbf{k}$ and $\omega$ and obtain the following equations. In the due process, we have neglected the 
second {and higher} order derivatives. The{nce the} corresponding equations are
\begin{subequations}
\begin{equation}
\begin{split}
\left(i\beta\alpha^2 - \frac{\alpha^4}{Re}\right)u(0)+2i\alpha_1\left(\frac{2\alpha^2}{Re} - 
i\beta\right)u'(0)\\+\frac{2i\alpha_3}{q}\zeta(0) + \frac{i}{4\pi} \left[(\textbf{B}_0\cdot\bm{\alpha})
(\alpha^2 B_x(0)-2i\alpha_1B_x^{\prime}(0))\right] \\= m_1,
\label{eq:dis_rel_1}
\end{split}
\end{equation}
\begin{eqnarray}
\begin{split}
i\alpha_3 \left(1-\frac{2}{q}\right)u(0)+\left(\frac{\alpha^2}{Re} - i\beta\right)\zeta(0)-\frac{2i\alpha_1}{Re}\zeta'(0)
\\-\frac{i}{4\pi} (\textbf{B}_0\cdot\bm{\alpha})\zeta_B(0)= m_2,
 \label{eq:dis_rel_2}
 \end{split}
\end{eqnarray}

\begin{eqnarray}
\begin{split}
 \left(\frac{\alpha^2}{Rm}- i\beta\right) B_x(0)-\frac{2i\alpha_1}{Rm} B_x^{\prime}(0)\\-i 
(\textbf{B}_0\cdot\bm{\alpha})u(0)= 0,
 \label{eq:dis_rel_3}
 \end{split}
\end{eqnarray}

\begin{eqnarray}
\begin{split}
 \left(\frac{\alpha^2}{Rm}- i\beta\right)\zeta_B(0)-\frac{2i\alpha_1}{Rm} \zeta_B^{\prime}(0)
\\-i (\textbf{B}_0\cdot\bm{\alpha})\zeta(0) -i\alpha_3 B_x(0)= 0.
 \label{eq:dis_rel_4}
 \end{split}
\end{eqnarray}
\end{subequations}

\subsection{Background magnetic field}
\label{bg_mag_field}
Before proceeding further, we have to {specify} the background magnetic field, $B_0$, like we {know} 
about 
$U_0$. The induction equation in dimensionless unit, in general, is 
\begin{eqnarray}
 \frac{\partial \bm{\BB}}{\partial t} = \bm{\nabla}\times(\textbf{V}\times\bm{\BB}) +\frac{1}{Rm}\nabla^2\bm{\BB},
 \label{eq:induction_eq}
\end{eqnarray}\\
where $\bm{\BB}$ and $\textbf{V}$ are arbitrary magnetic and velocity field vectors respectively. If we assume the 
background magnetic 
field to be constant over space and time, then equation (\ref{eq:induction_eq}) for background quantities becomes
\begin{equation}
\bm{\nabla}\times(\textbf{U}_0 \times \textbf{B}_0) = (\textbf{B}_0 \cdot \nabla)\textbf{U}_0 - (\textbf{U}_0 \cdot 
\nabla)\textbf{B}_0= 0.
\label{eq:induction_bg}
\end{equation}
For $\textbf{U}_0=(0,-x,0)$, the equation (\ref{eq:induction_bg}) becomes
\begin{equation}
-B_{0x}\textbf{j} + x\frac{\partial \textbf{B}_0}{\partial y} = 0,
\label{eq:bg_mag_eq_1}
\end{equation}
where $\textbf{j}$ is the unit vector along the $y$-direction in Cartesian coordinates. We, therefore, obtain that for 
constant background magnetic field, $B_{0x} = 0$, hence, $\textbf{B}_0 = (0, B_{0y}, B_{0z})$. {In this work, we 
shall be however focusing on the vertical magnetic field, i.e. $\textbf{B}_0=(0,0,B_{0z})$.} 

\section{The centrifugal instability and magnetorotational instability}
\label{sec:CI_MRI}
It is always interesting to obtain CI and MRI following our formalism. We, in fact, obtain the corresponding 
dispersion relations out of equations (\ref{eq:dis_rel_1}), (\ref{eq:dis_rel_2}), (\ref{eq:dis_rel_3}) and 
(\ref{eq:dis_rel_4}) omitting noise. Our primary interest is for 
vertical wave vector, i.e., $\bm{\alpha} = (0,0,\alpha_3)$. We also neglect the effect 
of viscosity for the time being. Eventually, {in this work itself,} we include all of them one by one in the due 
course of study. 
Considering all these, equations (\ref{eq:dis_rel_1}), (\ref{eq:dis_rel_2}), (\ref{eq:dis_rel_3}) and 
(\ref{eq:dis_rel_4}) 
become
\begin{eqnarray}
 \begin{split}
  i \beta \alpha_3^2 u(0) +\frac{2 i \alpha_3}{q}\zeta(0)+ \frac{i \alpha_3^3  B_{0z} B_{x}(0)}{4 \pi }=0,\\
  i \alpha_3\left(1-\frac{2}{q}\right)u(0) - i\beta \zeta(0) - \frac{i}{4\pi}B_{0z}\alpha_3\zeta_B(0) = 0,\\
  - i\beta B_x(0)-i B_{0z}\alpha_3 u(0)= 0,\\
  - i\beta\zeta_B(0)-i B_{0z}\alpha_3\zeta(0) -i\alpha_3 B_x(0)= 0.
  \label{eq:before_dis_rel_mri_ci}
 \end{split}
\end{eqnarray}
{For the nontrivial solutions of the above system of linear} equations,  we obtain the following dispersion relation 
\begin{equation}
\begin{split}
 16\pi^2\beta^4 - 8\pi\left[B_{0z}^2\alpha_3^2-\left(4\pi\left(\frac{1}{q}-\frac{2}{q^2}\right)\right)\right]\beta^2\\
 +B_{0z}^4\alpha_3^4-\frac{8\pi B_{0z}^2\alpha_3^2}{q} = 0.
 \label{eq:dis_rel_mri_ci_z}
 \end{split}
\end{equation}
For the centrifugal instability{\cite{Drazin_2004, Chandrasekhar_1961}}, we get rid of the magnetic field from the 
equation (\ref{eq:dis_rel_mri_ci_z}). The 
nontrivial solutions are 
\begin{equation}
 \beta = \pm\frac{\sqrt{2}}{q}\sqrt{2-q}.
 \label{eq:epicyclic_freq}
\end{equation}
It tells us that for $q > 2$, the flow becomes unstable. If we keep the magnetic field and get rid of the rotation 
parameter ($q$), we would obtain magneto-hydrodynamical (MHD) waves with frequency
\begin{equation}
 \beta = \pm\frac{B_{0z}\alpha_3}{2\sqrt{\pi}}.
 \label{eq:Alfven_freq}
\end{equation}
This waves are called Alfven wave with Alfven speed, $V_A = B_{0z}/2\sqrt{\pi}$. 

\begin{figure}   	
\includegraphics[width = \columnwidth]{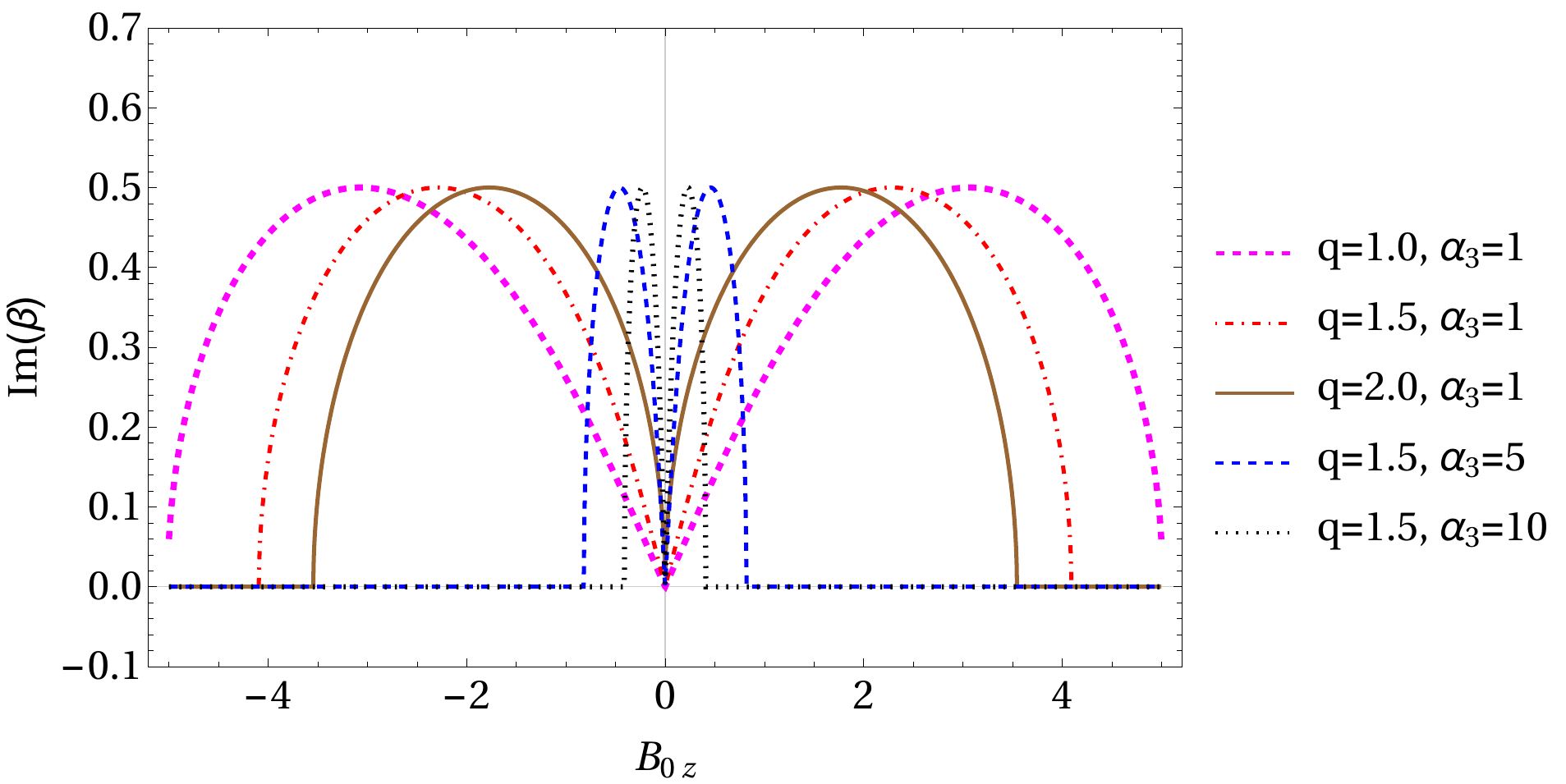}
\caption{Variation of $Im(\beta)$ of the unstable solution from equation (\ref{eq:dis_rel_mri_ci_z}) as a function of 
$B_{0z}$ for various combinations of $q$ and $\alpha_3$ {for inviscid flow}.}
\label{fig:MRI_diff_alpha_3}
\end{figure}

If we keep both rotation as well as the 
magnetic field, from the equation (\ref{eq:dis_rel_mri_ci_z}), we obtain that $q>0$ for instability, i.e., negative 
$\beta^2$ for an arbitrarily small $\alpha_3$. This is MRI. {E}quation 
(\ref{eq:dis_rel_mri_ci_z}) shows that $\beta$ has four solutions. However, among them only one provides instability. 
FIG.~\ref{fig:MRI_diff_alpha_3} shows the variation of 
$Im(\beta)$ as a function of $B_{0z}$ for various combination between $q = 1,\ 1.5,\ 2$ and $\alpha_3 = 1,\ 5,\ 10$. 
There we find that as the magnitude of 
$B_{0z}$ increases, first $Im(\beta)$ also increases and 
reaches a maximum value of $0.5$ \cite{Balbus_1998}, and then 
it decreases. 
It also says that for a fixed $\alpha_3$, as $q$  increase{s} the domain of $B_{0z}$ giving rise to positive 
$Im(\beta)$ decreases. On the other hand, for a fixed $q$ the increment of $\alpha_3$ decreases the window of $B_{0z}$, 
that gives rise to positive $Im(\beta)$. Between these effects {of $\alpha_3$ and $q$}, the increment of $\alpha_3$ 
for a fixed $q$ affects 
the flow more severely. 

\section{The effect of extra force}
\label{sec:effect_of_noise}
In the presence of an extra force, vertical magnetic field and the {vertical} wavevector, equations 
(\ref{eq:dis_rel_1}), 
(\ref{eq:dis_rel_2}), (\ref{eq:dis_rel_3}) and (\ref{eq:dis_rel_4}) become
\begin{eqnarray}
 \begin{split}
  i \beta \alpha_3^2 u(0) +\frac{2 i \alpha_3}{q}\zeta(0)+ \frac{i \alpha_3^3  B_{0z} B_{x}(0)}{4 \pi }=m_1,\\
  i \alpha_3\left(1-\frac{2}{q}\right)u(0) - i\beta \zeta(0) - \frac{i}{4\pi}B_{0z}\alpha_3\zeta_B(0) \\= m_2,\\
  - i\beta B_x(0)-i B_{0z}\alpha_3 u(0)= 0,\\
  - i\beta\zeta_B(0)-i B_{0z}\alpha_3\zeta(0) -i\alpha_3 B_x(0)= 0.
  \label{eq:before_dis_rel_hydro_noise}
 \end{split}
\end{eqnarray}
For simplicity, if we consider $m_1=m_2=m$ and write $u(0)$ as $u_0$, the dispersion relation from equation 
(\ref{eq:before_dis_rel_hydro_noise}) 
becomes
\begin{equation}
 \begin{split}
  \frac{m}{u_0} \left(4 \pi\alpha_3^2 B_{0z}^2 \beta-16\pi^2\beta^3\right) = \frac{8i\pi B_{0z}^2\alpha_3^4}{q} 
-i\alpha_3^6 B_{0z}^4\\ +\frac{32\pi^2 \alpha_3 \beta^2}{q}\left(\frac{m}{u_0}\right) +\frac{64i \pi^2 
\alpha_3^2\beta^2}{q^2} -\frac{32i\pi^2\alpha_3^2\beta^2}{q} \\+ 8i\pi\alpha_3^4\beta^2 B_{0z}^2 -16i\pi^2 \alpha_3^2 
\beta^4.
\label{eq:disp_with_noise_mag}
 \end{split}
\end{equation}

\subsection{Without magnetic field}
\label{sec:without_magenetic_field}
To extract the hydrodynamical effect out of equation (\ref{eq:disp_with_noise_mag}), we {set} $B_{0z} = 0$. Then the 
corresponding dispersion relation becomes
\begin{eqnarray}
 \begin{split}
 - i\alpha_3^2\beta^2 + \frac{m}{u_0}\beta+\frac{2 
\alpha_3}{q}\left(\frac{m}{u_0}\right)+\frac{4i\alpha_3^2}{q^2} 
\\-\frac{2i\alpha_3^2}{q} = 0.
\label{eq:disp_with_noise}
 \end{split}
\end{eqnarray}
The solutions of this quadratic equation are
\begin{equation}
\begin{split}
 &\beta =\\ &\frac{1}{2\alpha_3^2}\left[-i \left(\frac{m}{u_0}\right) \pm 
\sqrt{\frac{8\alpha_3^4}{q^2}(2-q)-\frac{m}{u_0}\left(\frac{m}{u_0}+\frac{8i\alpha_3^3}{q}\right)}\right].
\label{eq:hydro_sol_disp}
\end{split}
\end{equation}
If we compare equation (\ref{eq:hydro_sol_disp}) with equation (\ref{eq:epicyclic_freq}), we notice the effect of noise 
on the stability of the flow. Equation (\ref{eq:hydro_sol_disp}) has two solutions. However, only one of them gives us 
positive $Im(\beta)$ {as a function of} $m/u_0$, i.e., {only} that particular solution corresponds to instability 
in the flow. As we are interested in instability, we consider the 
first solution. In this case, the positive sign in front of the square root corresponds to the unstable solution.  
\begin{figure}		
\includegraphics[width = \columnwidth]{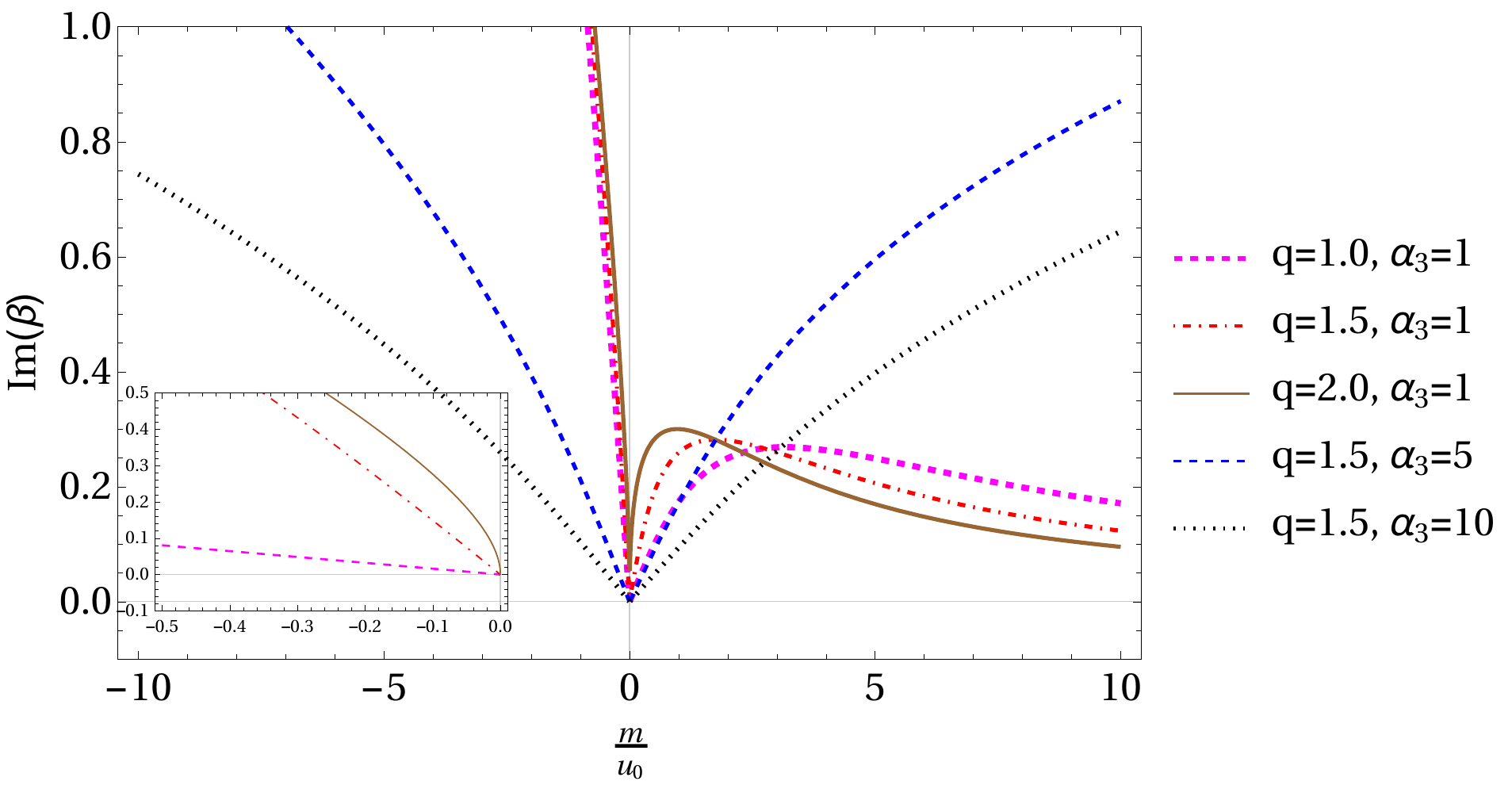}
\caption{Variation of $Im(\beta)$ {of} the unstable solution from equation (\ref{eq:disp_with_noise}) as a function 
of $m/u_0$ for various combinations of $q$ and $\alpha_3$ {for inviscid flow}.}
\label{fig:hydro_ins_diff_alpha_3_q}
\end{figure}

FIG.~\ref{fig:hydro_ins_diff_alpha_3_q} describes the variation of $Im(\beta)$ from equation (\ref{eq:hydro_sol_disp}) as 
a function of $m/u_0$ for various combinations of $q$ and $\alpha_3$. Here we notice that for positive $m/u_0$, when 
$\alpha_3$ has been kept fixed, $Im(\beta)$ reaches a maximum and then it decreases. The maximum decreases as $q$ 
decreases. Apart from that, the value of $m/u_0$, at which the maximum in $Im(\beta)$ occurs, increases as $q$ decreases. 
It, indeed, should be the case. Now for the negative $m/u_0$, for the same case, it is obvious from the inset that the 
{the} curve {of larger $q$} is steeper. It means that at each negative $m/u_0$, larger $q$ has larger 
$Im(\beta)$. This is what 
is expected. From equation (\ref{eq:epicyclic_freq}), {when the force is zero} $q=2$ gives the marginal stability and 
$q<2$ makes the flow stable 
and the stability increases as the $q$ becomes lesser than $2$. Lesser $q$, therefore, takes larger force to make the 
flow unstable. For a fixed $q$, as $\alpha_3$ increase{s,} the curves become less steeper in most of the cases, hence 
at 
each $m/u_0$ lesser $\alpha_3$ has larger growth rate [$Im(\beta)$]. This nature of the the growth rate for 
negative $m/u_0$ is quite obvious from the equation (\ref{eq:hydro_sol_disp}). However, for the positive $m/u_0$, we 
notice that there will be a competition between the terms with and with without square root in equation 
(\ref{eq:hydro_sol_disp}). As a result, for $\alpha_3 = 1$ and $q = 1.5$, we see that the growth rate increases as 
$m/u_0$ increases up to certain value of $m/u_0$. Beyond that certain $m/u_0$, the growth rate decreases. 

\subsection{With magnetic field}
\label{sec:With_magnetic_field}
The equation (\ref{eq:disp_with_noise_mag}) is a quartic equation. Here we look for the imaginary part of each root. 
Among the four solutions:
\begin{itemize}
 \item  the first and the fourth solutions together give us the solution given by equation 
(\ref{eq:hydro_sol_disp}) when magnetic field is switched off,
\item {t}he third solution gives us the MRI modes when noise is not there in the flow, 
\item {t}he second solution gives us the stable modes corresponding to the MRI modes if the noise is made 
zero.
\end{itemize} 
We will begin with the `third {solution'} as it reduces to the usual MRI mode shown in 
FIG.~\ref{fig:MRI_diff_alpha_3}, when 
there is no noise in the system. 
\subsubsection{The {`third solution'}}
\label{sec:the_third_sol}
\begin{figure}		
\includegraphics[width = \columnwidth]{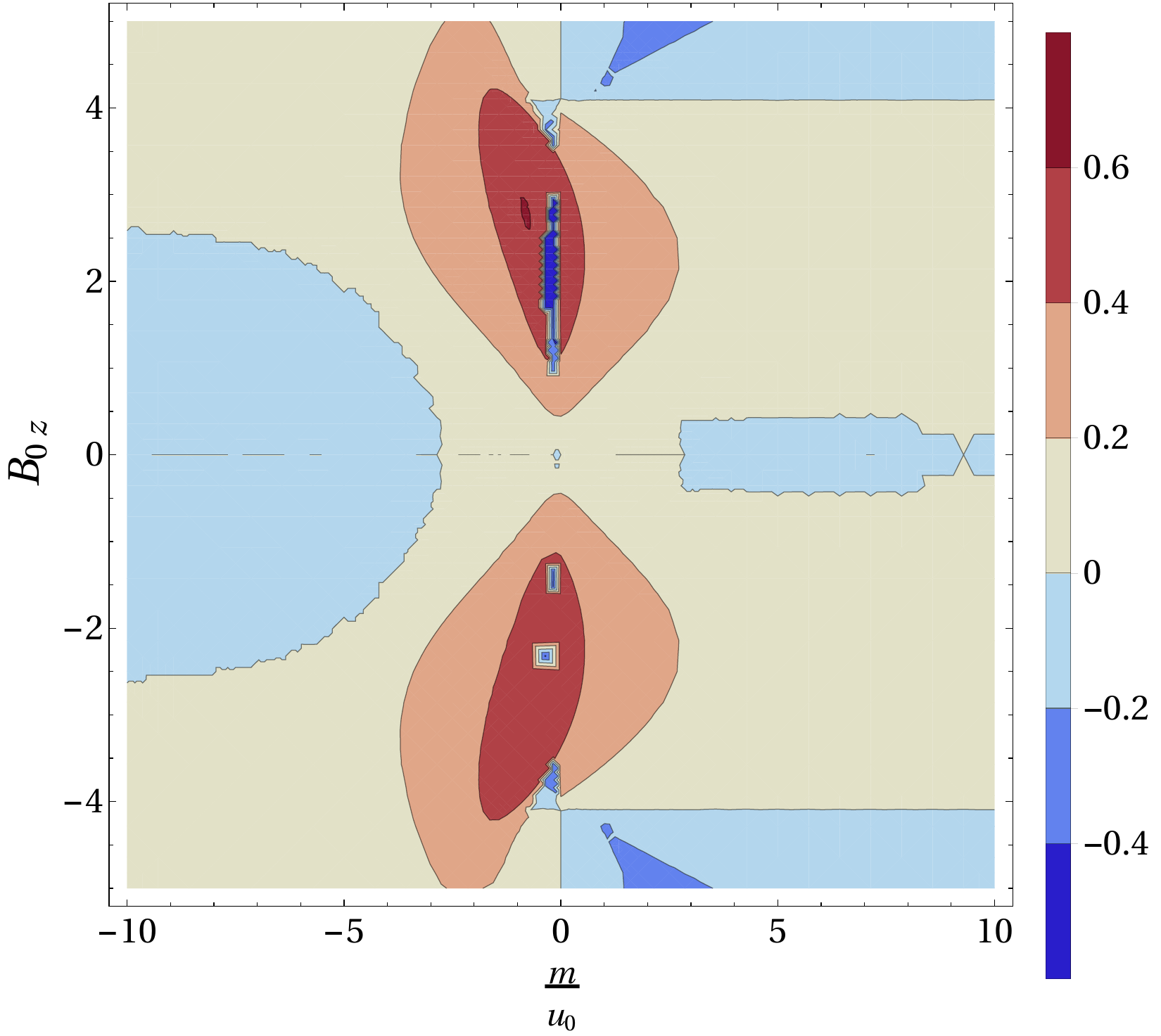}
\caption{Variation of $Im(\beta)$ of the {`third solution'} of equation (\ref{eq:disp_with_noise_mag}) {as described in 
\S\ref{sec:With_magnetic_field}} as a function of $m/u_0$ and $B_{0z}$ for $q = 1.5$ and $\alpha_3 = 1$ {for inviscid 
flow}.}
\label{fig:beta3_alpha3_1_kep_small_p_B3}
\end{figure}

\begin{figure}		
\includegraphics[width = \columnwidth]{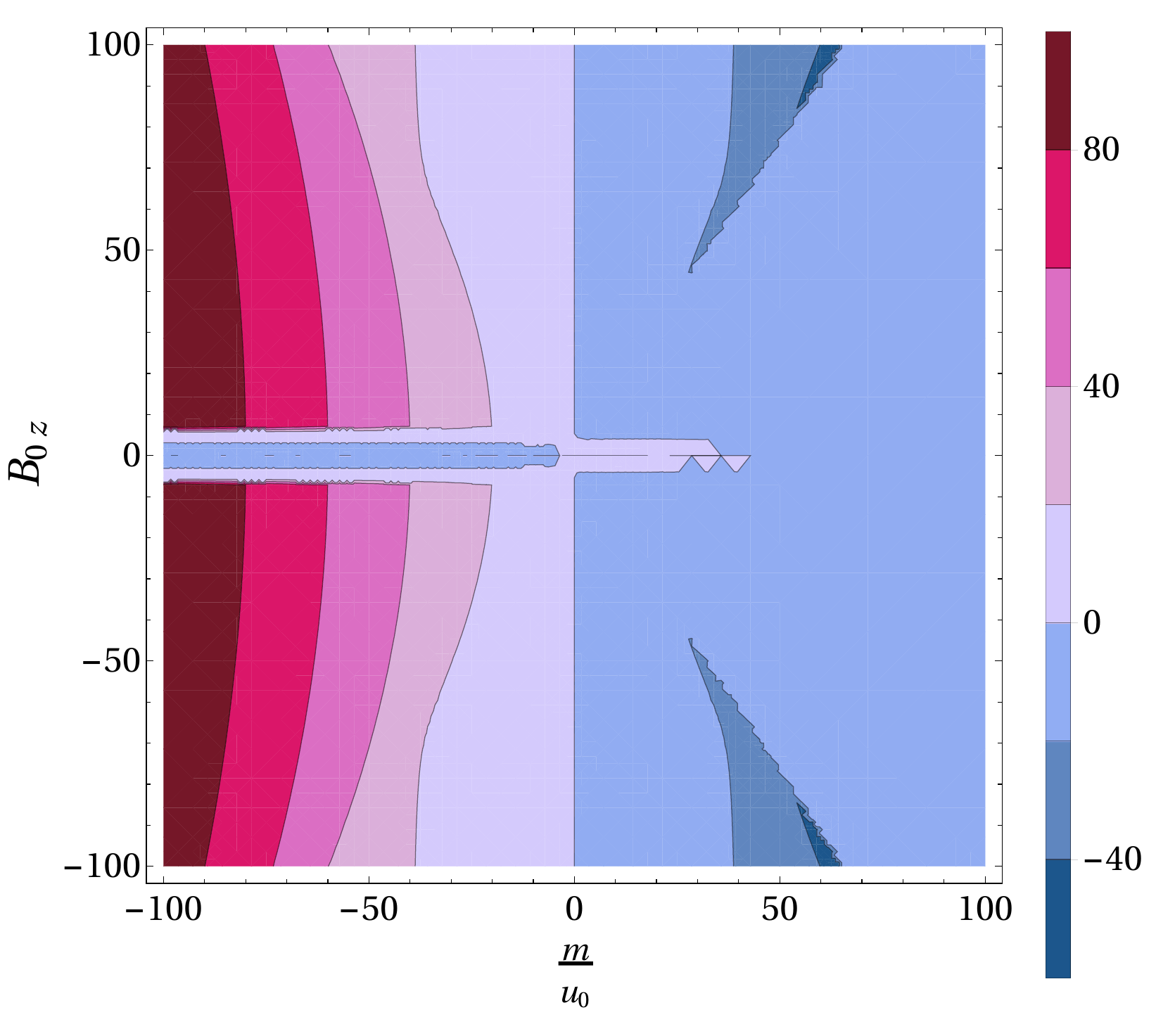}
\caption{Variation of $Im(\beta)$ of the {`third solution'} of equation (\ref{eq:disp_with_noise_mag}) {as described in 
\S\ref{sec:With_magnetic_field}} as a function of $m/u_0$ and $B_{0z}$ in the larger domain for $q = 1.5$ and $\alpha_3 = 
1$ {for inviscid flow}.}
\label{fig:beta3_alpha3_1_kep_large_p_B3}
\end{figure}

\begin{figure}		
\includegraphics[width = \columnwidth]{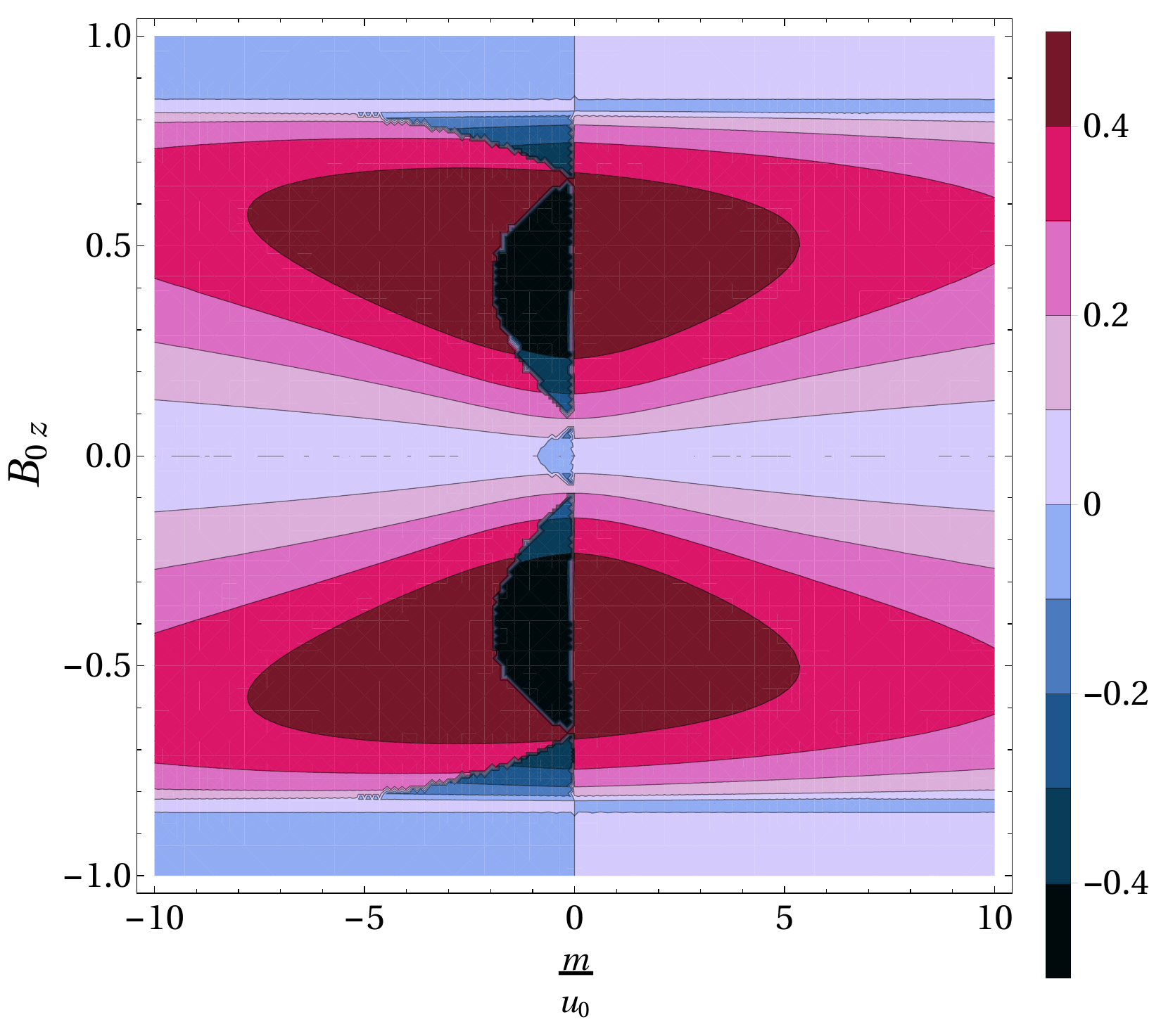}
\caption{Variation of $Im(\beta)$ of the {`third solution'} of equation (\ref{eq:disp_with_noise_mag}) {as described in 
\S\ref{sec:With_magnetic_field}} as a function of $m/u_0$ and $B_{0z}$ in for $q = 1.5$ and $\alpha_3 = 5$ {for 
inviscid flow}.}
\label{fig:3rd_sol_noise_alpha_5_p_10_B_1}
\end{figure}

\begin{figure}		
\includegraphics[width = \columnwidth]{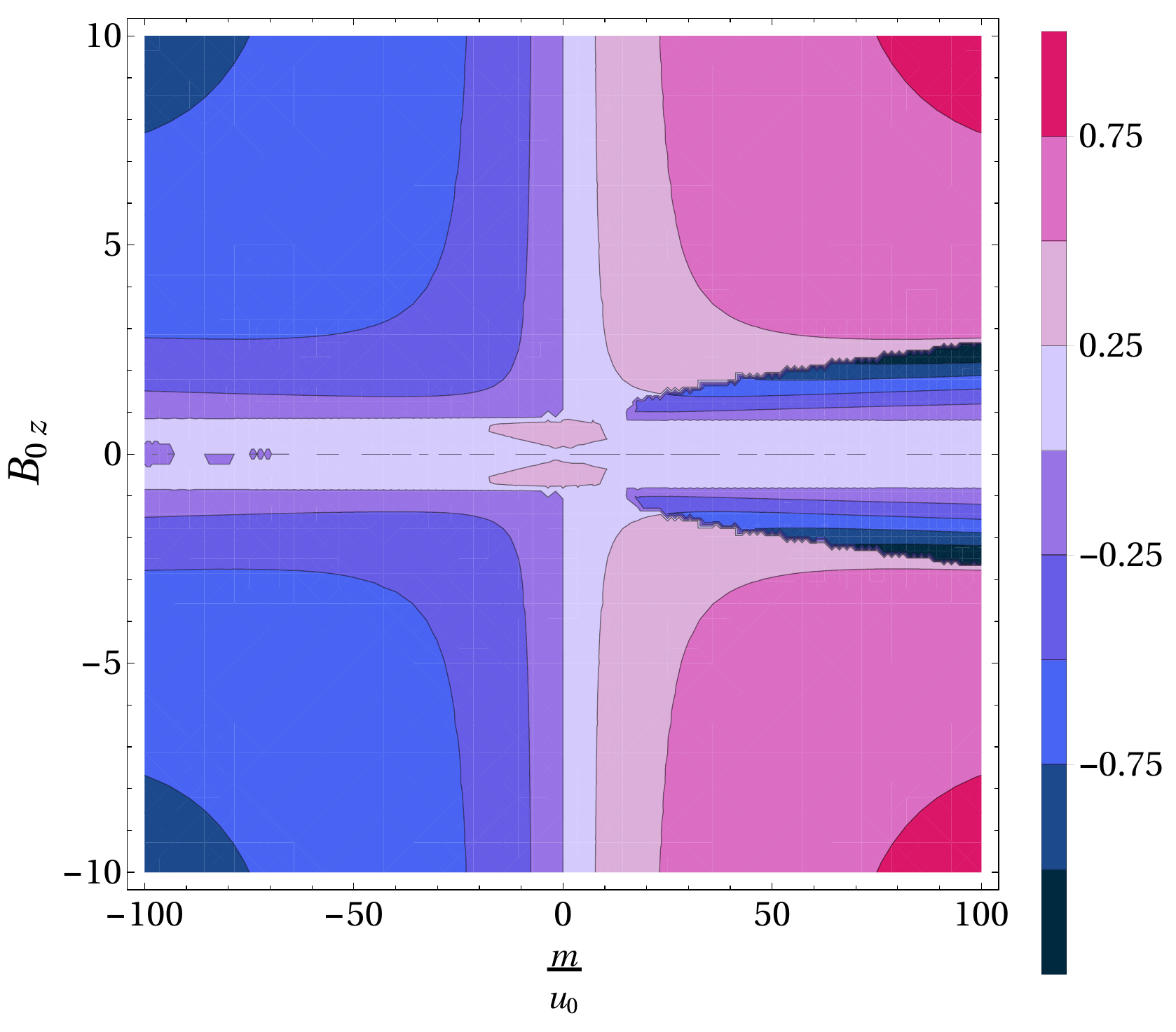}
\caption{Variation of $Im(\beta)$ of the {`third solution'} of equation (\ref{eq:disp_with_noise_mag}) {as described in 
\S\ref{sec:With_magnetic_field}} as a function of $m/u_0$ and $B_{0z}$ in the larger domain for $q = 1.5$ and $\alpha_3 = 
5$ {for inviscid flow}.}
\label{fig:3rd_sol_noise_alpha_5_p_100_B_10}
\end{figure}

\begin{figure}		
\includegraphics[width = \columnwidth]{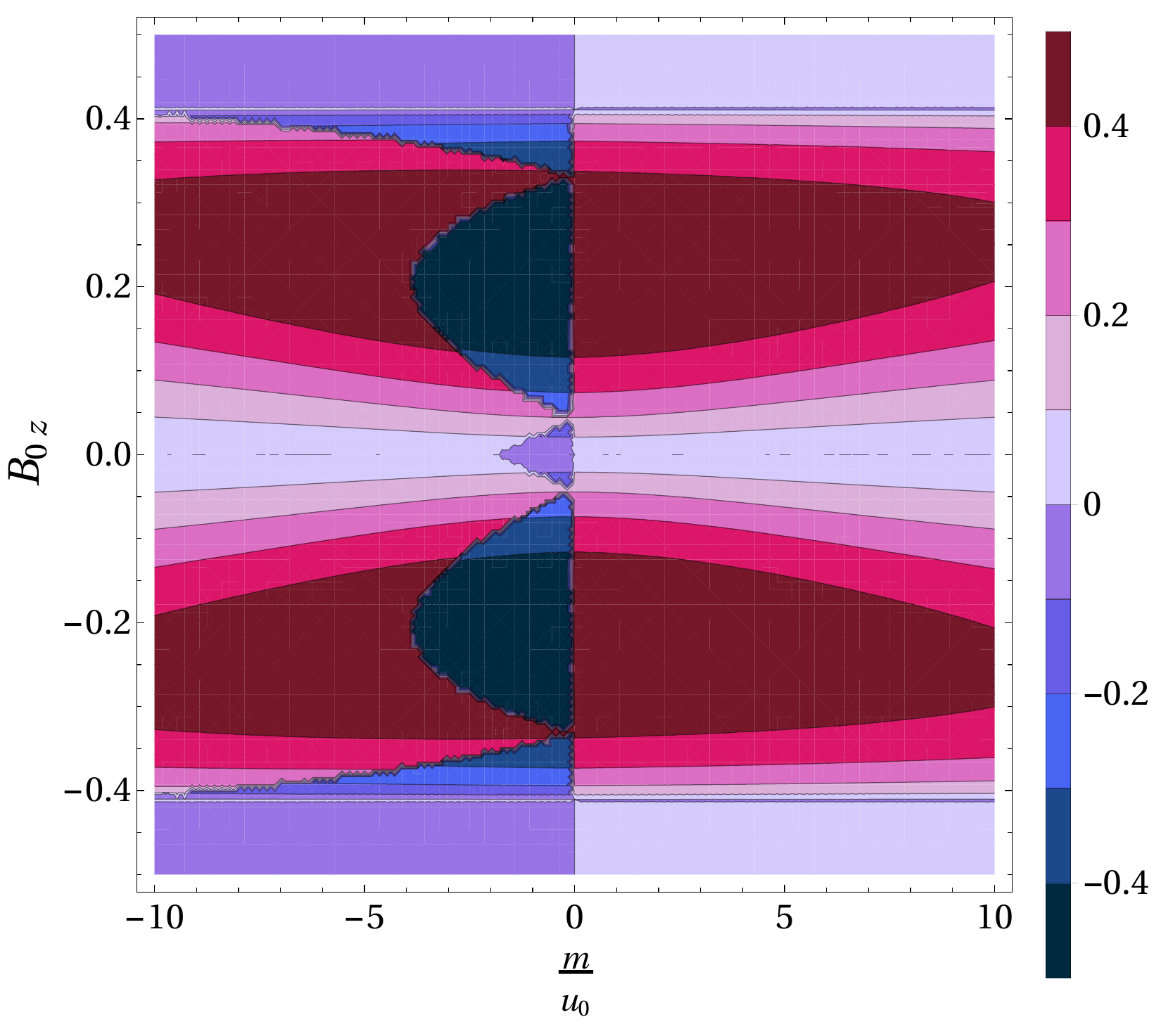}
\caption{Variation of $Im(\beta)$ of the {`third solution'} of equation (\ref{eq:disp_with_noise_mag}) {as described in 
\S\ref{sec:With_magnetic_field}} as a function of $m/u_0$ and $B_{0z}$ for $q = 1.5$ and $\alpha_3 = 10$ {for inviscid 
flow}.}
\label{fig:3rd_sol_noise_alpha_10_p_10_B_point5} 
\end{figure}

\begin{figure}		
\includegraphics[width = \columnwidth]{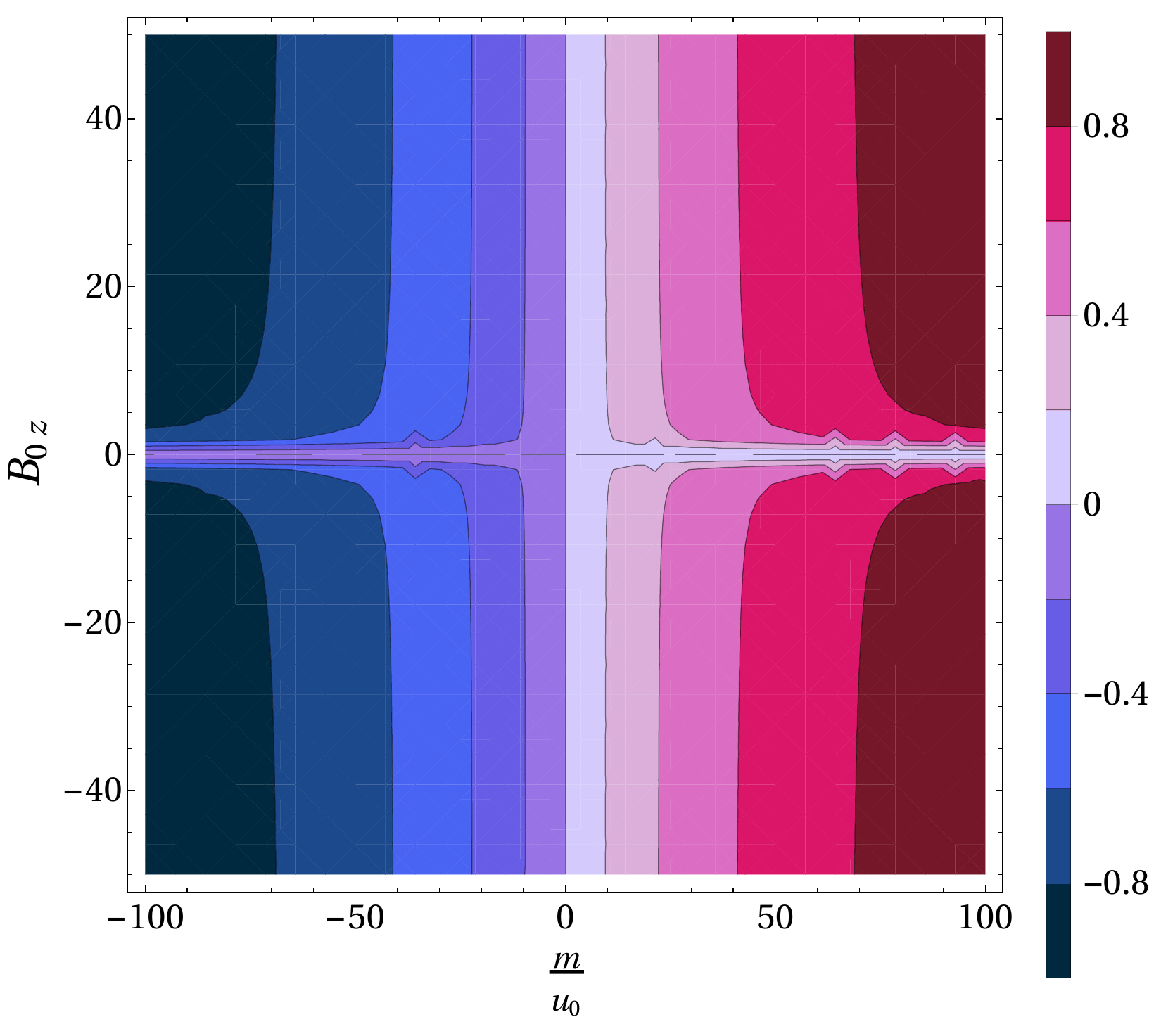}
\caption{Variation of $Im(\beta)$ of the {`third solution'} of equation (\ref{eq:disp_with_noise_mag}) {as described in 
\S\ref{sec:With_magnetic_field}} as a function of $m/u_0$ and $B_{0z}$ in the larger domain for $q = 1.5$ and 
$\alpha_3 = 10$ {for inviscid flow}.}
\label{fig:3rd_sol_noise_alpha_10_p_100_B_50}
\end{figure}

FIG.~\ref{fig:beta3_alpha3_1_kep_small_p_B3} describes the variation of $Im(\beta)$ as a function of $m/u_0$ and $B_{0z}$ 
for $q = 1.5$ and $\alpha_3 = 1$. From FIG.~\ref{fig:MRI_diff_alpha_3} it is clear that for $q = 1.5$ and $\alpha_3 = 1$, 
the domain of $B_{0z}$ that gives rise to positive $Im(\beta)$ is {about} within {$-4$ to $+4$}. To see what 
happens in that 
region of $B_{0z}$, we fix the range of the {vertical} axis in FIG.~\ref{fig:beta3_alpha3_1_kep_small_p_B3} within 
{$-5$ to $+5$}. We 
notice that the growth rate is mostly dominated by MRI in that region of the magnetic field. Around $m/u_0 = 0$, as 
the magnitude of $B_{0z}$ increases, $Im(\beta)$ increases initially, reaches a maximum, and then again decays. Also, 
beyond a small region {around} $|m/u_0| = 0$, the flow is stable within the 
domain of $m/u_0$ as shown in the 
FIG.~\ref{fig:beta3_alpha3_1_kep_small_p_B3}. However, the maximum growth rate shown in this figure is greater than 
MRI growth rate shown in FIG.~\ref{fig:MRI_diff_alpha_3}. This could be due to the additional effect from the 
hydrodynamic instability because of the presence of noise in the flow. 

Now, if we increase the domain of $m/u_0$ and $B_{0z}$, we notice that the flow is {mostly} stable for positive 
$m/u_0$, but it is 
{mostly} unstable for negative $m/u_0$ as indicated by the FIG.~\ref{fig:beta3_alpha3_1_kep_large_p_B3}, where the 
variation of 
$Im(\beta)$ is shown as a function of $m/u_0$ and $B_{0z}$ in the larger domain. As the magnitude of $m/u_0$ increases, 
we notice that the growth rate increases. However, the {positive} growth rate becomes almost independent of the 
magnetic field at 
large $m/u_0$. That is to say, at the larger magnetic field where there is no growth due to MRI, the flow is mostly 
governed by the noise with huge growth rates. 

Now, we increase $\alpha_3$ to $5$. FIG.~\ref{fig:3rd_sol_noise_alpha_5_p_10_B_1} shows the variation of $Im(\beta)$ as a 
function of $m/u_0$ and $B_{0z}$ for $q = 1.5$ and $\alpha_3= 5$. {F}rom FIG.~\ref{fig:MRI_diff_alpha_3} we 
{already have obtained} that for $q = 1.5$ and $\alpha_3= 5$, the window of $B_{0z}$ that gives rise to positive 
$Im(\beta)$ is {about} 
within {$-1$ to $+1$}. We, therefore, keep the range of the {vertical} axis 
in FIG.~\ref{fig:3rd_sol_noise_alpha_5_p_10_B_1} within 
$\pm1$ so that we can compare between MRI and instability due to noise. There we notice that the growth rate is 
mostly governed by the MRI and the the structure of the growth rate, i.e., how $Im(\beta)$ varies with $B_{0z}$ at a 
particular $m/u_0$, is also MRI-like. If we compare between the FIGs.~ \ref{fig:beta3_alpha3_1_kep_small_p_B3} 
and \ref{fig:3rd_sol_noise_alpha_5_p_10_B_1}, we observe that the range of $|m/u_0|$, that gives rise to a positive 
growth rate, increases as $\alpha_3$ increases. 

Now let us see what happens to the growth rates, when we extend the domain 
of $m/u_0$ and $B_{0z}$ through the FIG.~\ref{fig:3rd_sol_noise_alpha_5_p_100_B_10}, where the variation of $Im(\beta)$
has been shown {as} a function of $m/u_0$ and $B_{0z}$ for $q = 1.5$ and $\alpha_3 = 5$. There we notice that for 
negative 
$m/u_0$, the flow is stable for any magnetic field. On the other hand, for positive $m/u_0$, the flow is {mostly} 
unstable. 
However, there is a window of {$-3 \lesssim B_{0z} \lesssim +3$}, where the flow could be stable 
depending on $m/u_0$. Here noise has the 
stabilizing effect 
because 
if we fix the magnetic field at any value within $\pm3$, we notice that the flow suddenly becomes stable as we 
increase $m/u_0$. However, in the unstable region, as the noise increases, the growth rate increases, and ultimately at 
large noise, the instability, due to the noise, takes over the flow. 
{{Thus,} the regime of sterile magnetic field not 
leading to MRI, i.e. the strong field, turns out to be active in order to reveal instability in the presence of noise.}

Now, we further increase $\alpha_3$ to $10$, which is in fact very much applicable to the accretion flow 
\cite{Ghosh_2021_a}. 
FIG.~\ref{fig:3rd_sol_noise_alpha_10_p_10_B_point5} describes the variation of $Im(\beta)$ as a function of $m/u_0$ and 
$B_{0z}$ for $q = 1.5$ and $\alpha_3 = 10$. We bound the {vertical} axis within $\pm0.5$ in this figure as 
FIG.~\ref{fig:MRI_diff_alpha_3} {shows} that the range of $B_{0z}$ that gives rise to a positive growth rate is 
{within around} 
$\pm 0.5$. Here again, we notice that in this domain of magnetic field, the growth rate is mostly dominated by MRI. 
However, in comparison with the other two $\alpha_3$, for $\alpha_3 = 10$, the range of $m/u_0$, which gives rise to a  
positive growth rate, is larger. Nevertheless, for the larger magnetic field, where MRI is inactive, the flow is again
dominated by the instability due to the noise, particularly {for} positive $m/u_0$. This phenomenon is well-depicted 
in the 
FIG.~\ref{fig:3rd_sol_noise_alpha_10_p_100_B_50} which shows the variation of $Im(\beta)$ as a function of $m/u_0$ and 
$B_{0z}$ in the larger domain for $q = 1.5$ and $\alpha_3 = 10$. 

\subsubsection{The {`first solution'}}
\label{sec:the_first_sol}
\begin{figure}		
\includegraphics[width = \columnwidth]{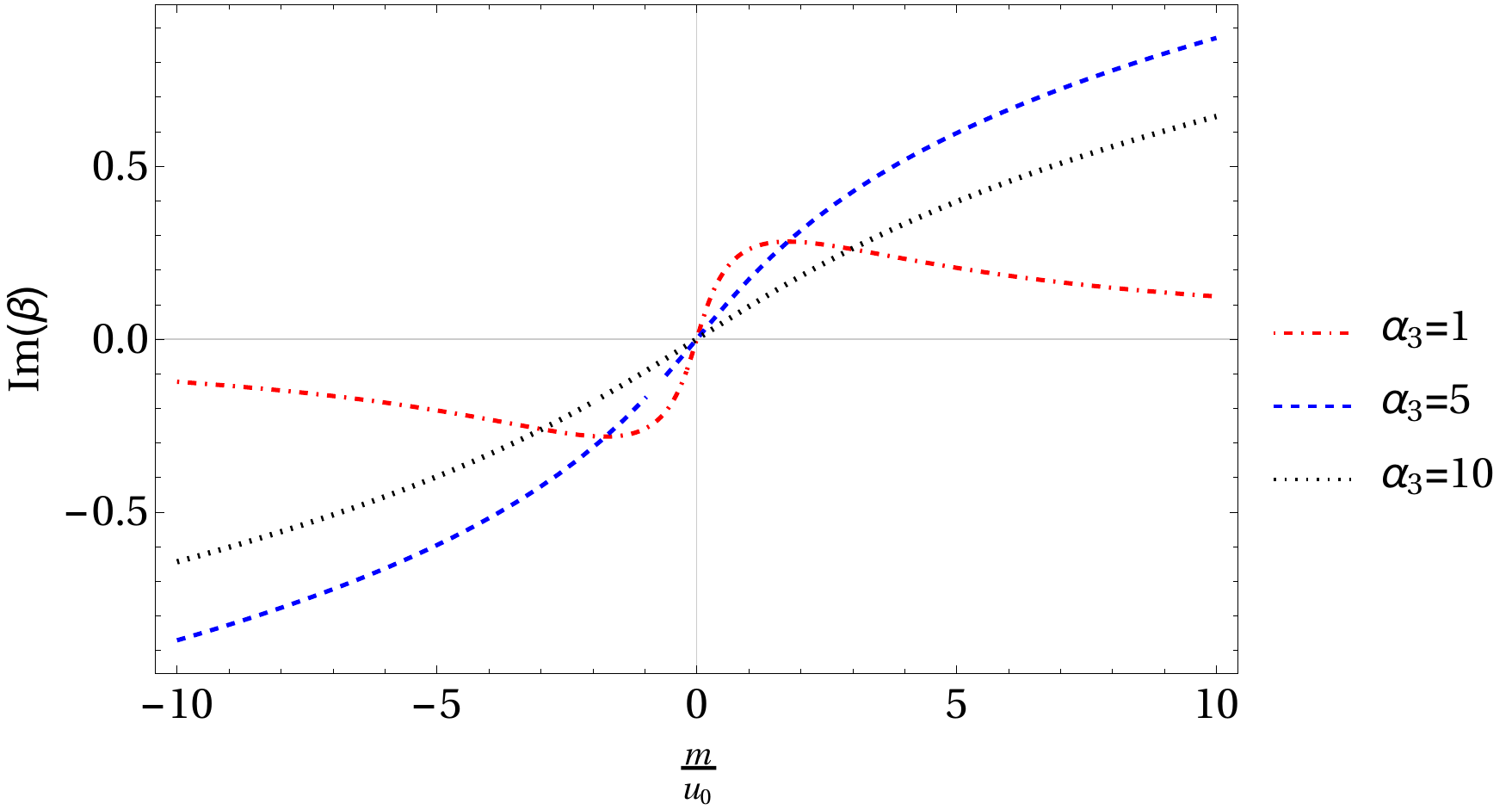}
\caption{Variation of $Im(\beta)$ of the {`first solution'} of equation (\ref{eq:disp_with_noise_mag}) {as described in 
\S\ref{sec:With_magnetic_field}} as a function of $m/u_0$ for $q = 1.5$ and {three different $\alpha_3$} 
when $B_{0z} = 0$ {for inviscid flow}.}
\label{fig:sol1_hydro_inst_diff_alpha_3}
\end{figure}

\begin{figure}		
\includegraphics[width = \columnwidth]{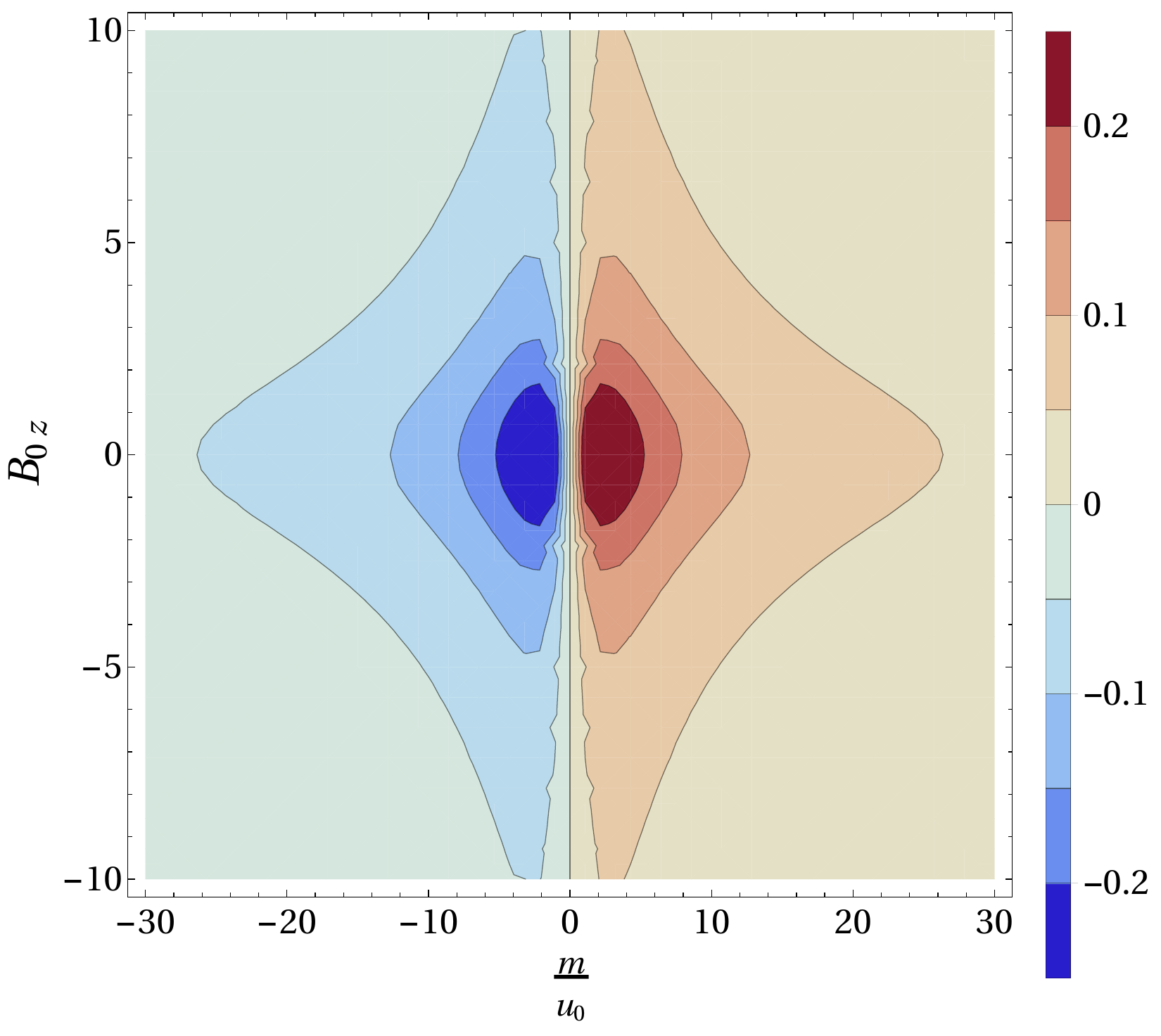}
\caption{Variation of $Im(\beta)$ of the {`first solution'} of equation (\ref{eq:disp_with_noise_mag}) {as described in 
\S\ref{sec:With_magnetic_field}} as a function of $m/u_0$ and $B_{0z}$ for $q = 1.5$ and $\alpha_3 = 1$ {for inviscid 
flow}.}
\label{fig:1st_sol_noise_alpha_1_p_30_B_10}
\end{figure}

\begin{figure}		
\includegraphics[width = \columnwidth]{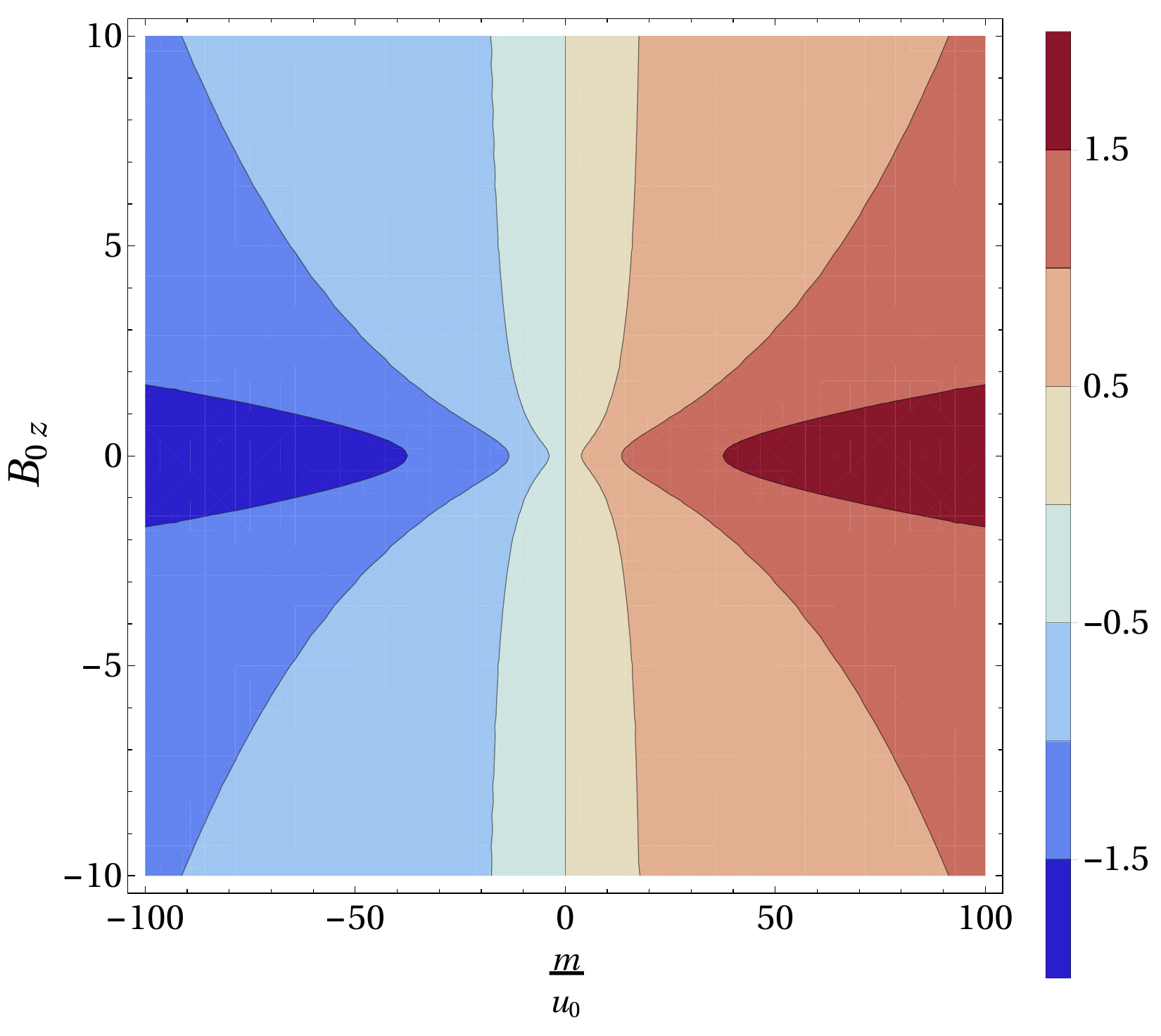}
\caption{Variation of $Im(\beta)$ of the {`first solution'} of equation (\ref{eq:disp_with_noise_mag}) {as described in 
\S\ref{sec:With_magnetic_field}} as a function of $m/u_0$ and $B_{0z}$ for $q = 1.5$ and $\alpha_3 = 5$ {for inviscid 
flow}.}
\label{fig:1st_sol_noise_alpha_5_p_100_B_10}
\end{figure}

\begin{figure}		
\includegraphics[width = \columnwidth]{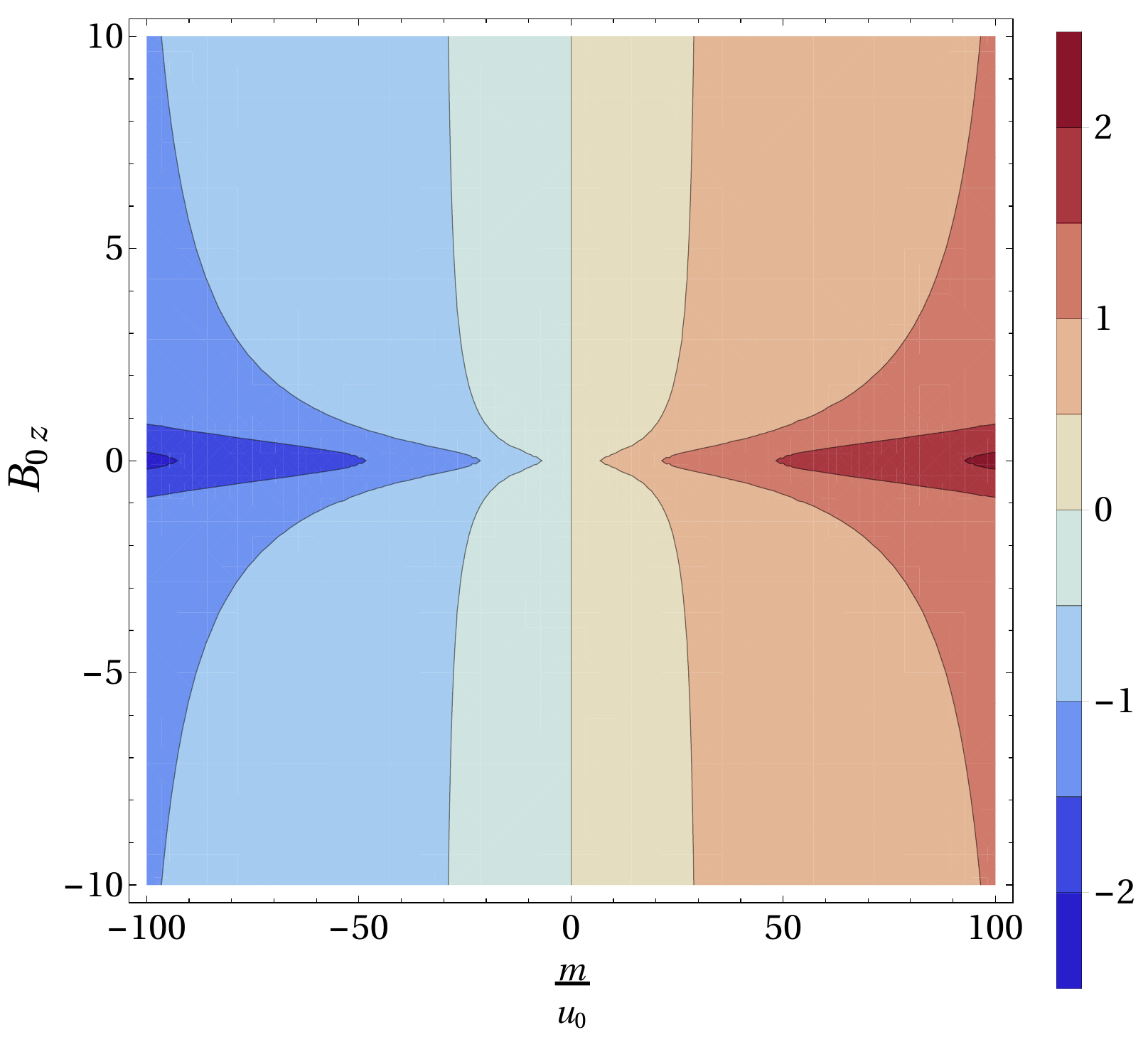}
\caption{Variation of $Im(\beta)$ of the {`first solution'} of equation (\ref{eq:disp_with_noise_mag}) {as described in 
\S\ref{sec:With_magnetic_field}} as a function of $m/u_0$ and $B_{0z}$ for $q = 1.5$ and $\alpha_3 = 10$ {for inviscid 
flow}.}
\label{fig:1st_sol_noise_alpha_10_p_100_B_10}
\end{figure}

The {`first solution'} of $Im(\beta)$ from equation (\ref{eq:disp_with_noise_mag}) {as 
described in \S\ref{sec:With_magnetic_field}} is shown in 
FIG.~\ref{fig:sol1_hydro_inst_diff_alpha_3} for $B_{0z} = 0$ and three different $\alpha_3$. There we notice that the 
variation of $Im(\beta)$ for positive $m/u_0$ {is} the same {as} of FIG.~\ref{fig:hydro_ins_diff_alpha_3_q} for 
positive $m/u_0$. We see that up to certain {positive} $m/u_0$, the growth rates corresponding to the smaller 
$\alpha_3$ {increase} steep{ly}. However, beyond that $m/u_0$, the growth rate decreases. The similar 
{feature} is depicted in FIG.~\ref{fig:1st_sol_noise_alpha_1_p_30_B_10} where the variation of the growth rate is 
shown a function of $m/u_0$ and $B_{0z}$ for $\alpha_3 = 1$. However, for a particular $m/u_0$, we see that the growth 
rate decreases with the increase of the magnitude of $B_{0z}$. Hence, in this case, the magnetic field {affects} the flow 
destructively with the effect of noise. 

FIGs.~\ref{fig:1st_sol_noise_alpha_5_p_100_B_10} and \ref{fig:1st_sol_noise_alpha_10_p_100_B_10} show the variation of 
$Im(\beta)$ of the {`first solution'} of equation (\ref{eq:disp_with_noise_mag}) {as described in 
\S\ref{sec:With_magnetic_field}} as a function of $m/u_0$ and $B_{0z}$ for $q = 1.5$, for $\alpha_3 =5 $ and $10$, 
respectively. There we notice that the growth rates for these two cases are similar to that in the 
FIG.~\ref{fig:1st_sol_noise_alpha_1_p_30_B_10}. Within a very small domain of magnetic field, for a fixed $m/u_0$, the 
growth rate decreases with the increase in the magnitude of the magnetic field. Beyond that domain of $B_{0z}$, the 
growth rate is almost independent of magnetic field. However, the domain of magnetic field, in which the growth rate 
depends on the magnetic field, shrinks in size as $\alpha_3$ increase{s} to 10 from 5. {It could 
plausibly be
because nonzero noise effectively brings in MRI-like features, i.e. the decrease in size of the MRI active magnetic field 
regime due 
to the increase in wavevector, for a particular region of magnetic field.}

{The maximum growth rate in FIG.~\ref{fig:1st_sol_noise_alpha_10_p_100_B_10}, is relatively larger than that in  
FIG.~\ref{fig:3rd_sol_noise_alpha_10_p_100_B_50}. Note that both of these figures have the same $\alpha_3$ and the same 
range 
of $m/u_0$. While FIG.~\ref{fig:3rd_sol_noise_alpha_10_p_100_B_50} represents the `third solution' for 
a given set of parameters, FIG.~\ref{fig:1st_sol_noise_alpha_10_p_100_B_10} represents the `first 
solution' for the concerned parameters. The latter one, being solely hydrodynamical, does not have any MRI counterpart. 
On the contrary, the former one does reduce to the MRI solution at the vanishing noise. }

\subsubsection{The {`second solution'}}
\label{sec:the_second_sol}
\begin{figure}		
\includegraphics[width = \columnwidth]{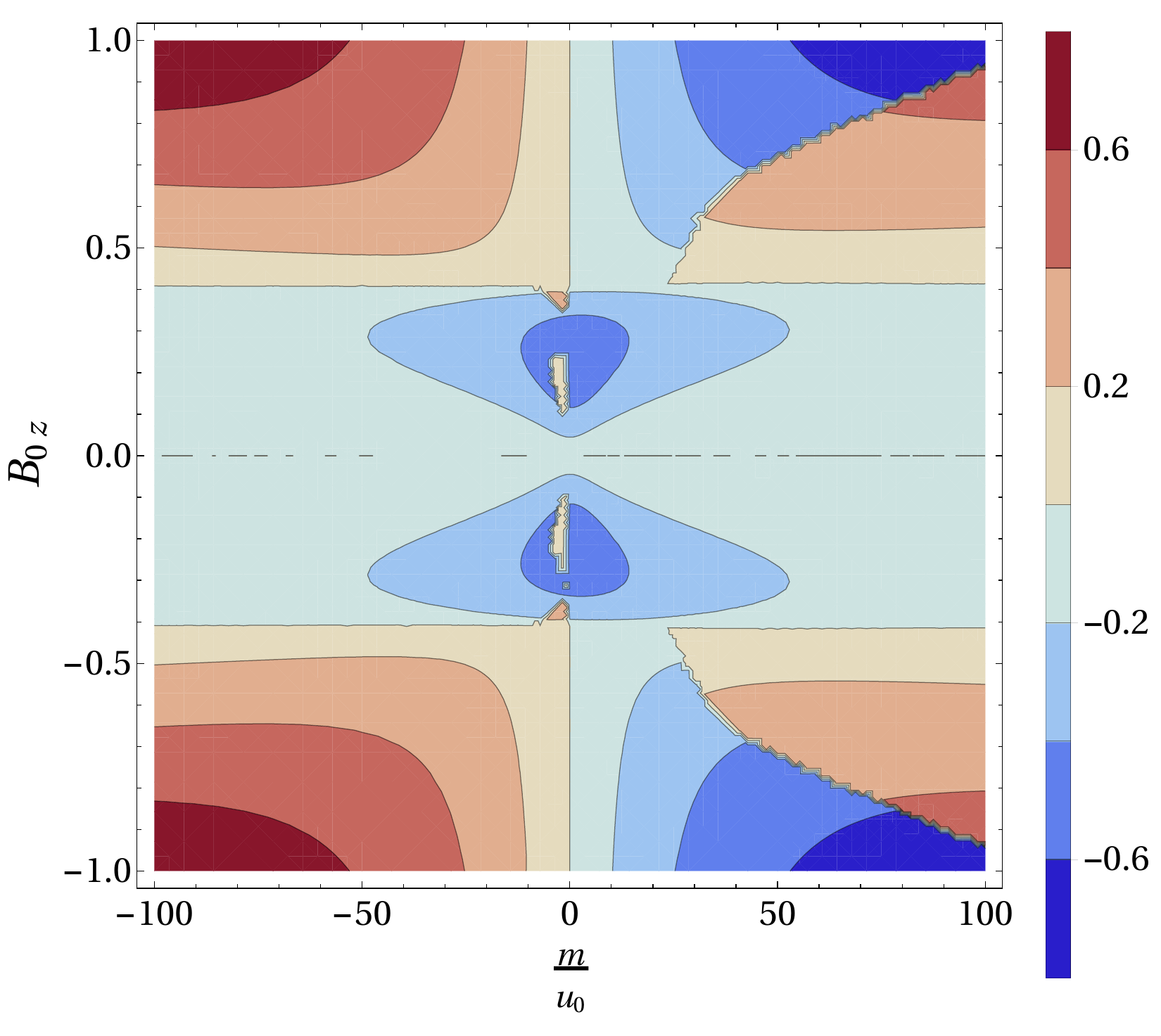}
\caption{Variation of $Im(\beta)$ of the {`second solution'} of equation (\ref{eq:disp_with_noise_mag}) {as described in 
\S\ref{sec:With_magnetic_field}} as a function of $m/u_0$ and $B_{0z}$ for $q = 1.5$ and $\alpha_3 = 10$ {for inviscid 
flow}.}
\label{fig:2nd_sol_noise_alpha_10_p_100_B_1}
\end{figure}

\begin{figure}		
\includegraphics[width = \columnwidth]{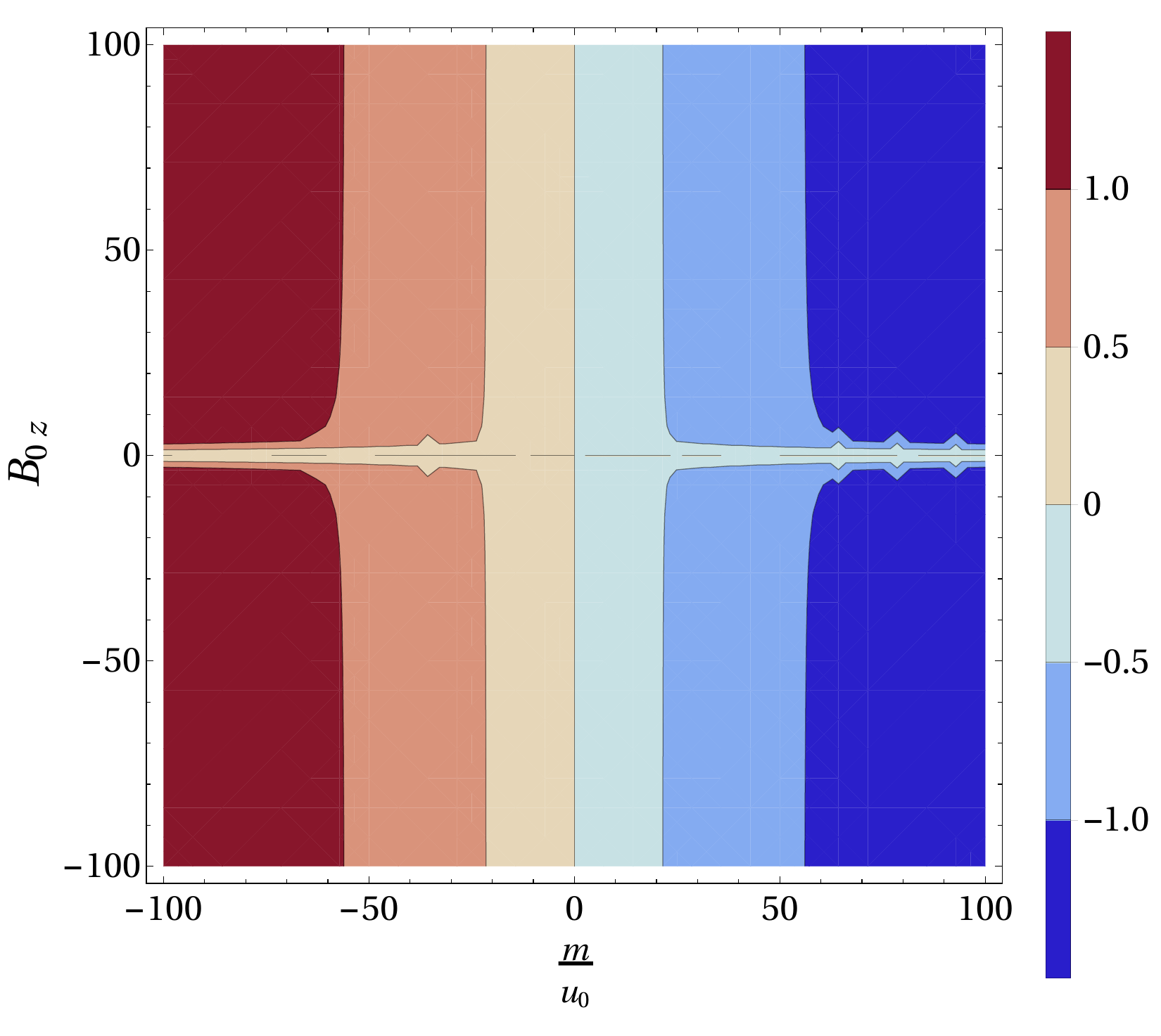}
\caption{Variation of $Im(\beta)$ of the {`second solution'} of equation (\ref{eq:disp_with_noise_mag}) {as described in 
\S\ref{sec:With_magnetic_field}} as a function of $m/u_0$ and $B_{0z}$ in the larger domain for $q = 1.5$ and $\alpha_3 = 
10$ {for inviscid flow}.}
\label{fig:2nd_sol_noise_alpha_10_p_100_B_100}
\end{figure}

The {`second solution'} of equation (\ref{eq:disp_with_noise_mag}) is stable version of the third solution of the 
same 
equation if there is no noise in the flow. FIG.~\ref{fig:2nd_sol_noise_alpha_10_p_100_B_1} shows the variation of 
$Im(\beta)$ of the second solution of equation (\ref{eq:disp_with_noise_mag}) as a function of 
$m/u_0$ and $B_{0z}$ for $q = 1.5$ and $\alpha_3 = 10$. FIG.~\ref{fig:2nd_sol_noise_alpha_10_p_100_B_100} also depicts 
the same but in the larger domain of $m/u_0$ and $B_{0z}$. We notice that FIG.~\ref{fig:2nd_sol_noise_alpha_10_p_100_B_1} 
is almost complementary to the FIG.~\ref{fig:3rd_sol_noise_alpha_5_p_100_B_10}. Within a certain domain of $B_{0z}$, the 
flow is stable, and the stability arises due to the domination of the magnetic field over the effect of noise. However, 
beyond that particular domain of $B_{0z}$, the flow becomes unstable due the effect of noise. For negative $m/u_0$, we 
notice that beyond certain magnetic field, the growth rate increases as the magnitude of $m/u_0$ increases. In this 
region, the growth rate increases with the increase of magnitude of both $m/u_0$ and $B_{0z}$. However, at the larger 
magnetic field, the flow is almost controlled by the noise particularly with negative mean. We are also familiar with 
similar invasion of noise but with positive mean in case of the third solution {described in} \ref{sec:the_third_sol}.

\subsubsection{The {`fourth solution`}}
\label{sec:the_fourth_sol}
\begin{figure}		
\includegraphics[width = \columnwidth]{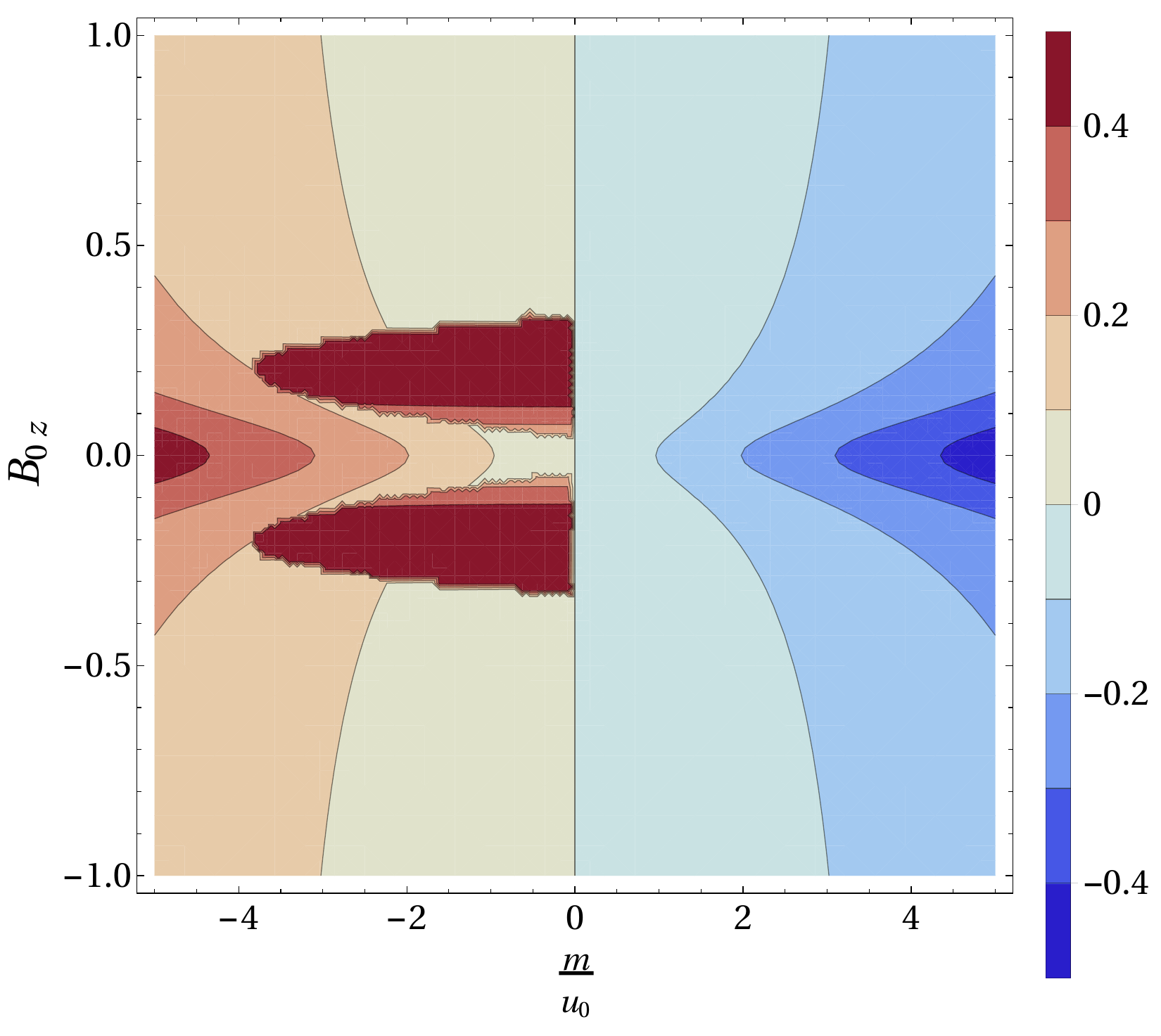}
\caption{Variation of $Im(\beta)$ of the {`fourth solution'} of equation (\ref{eq:disp_with_noise_mag}) {as described in 
\S\ref{sec:With_magnetic_field}} as a function of $m/u_0$ and $B_{0z}$ for $q = 1.5$ and $\alpha_3 = 10$ {for inviscid 
flow}.}
\label{fig:4th_sol_noise_alpha_10_p_5_B_1}
\end{figure}

\begin{figure}		
\includegraphics[width = \columnwidth]{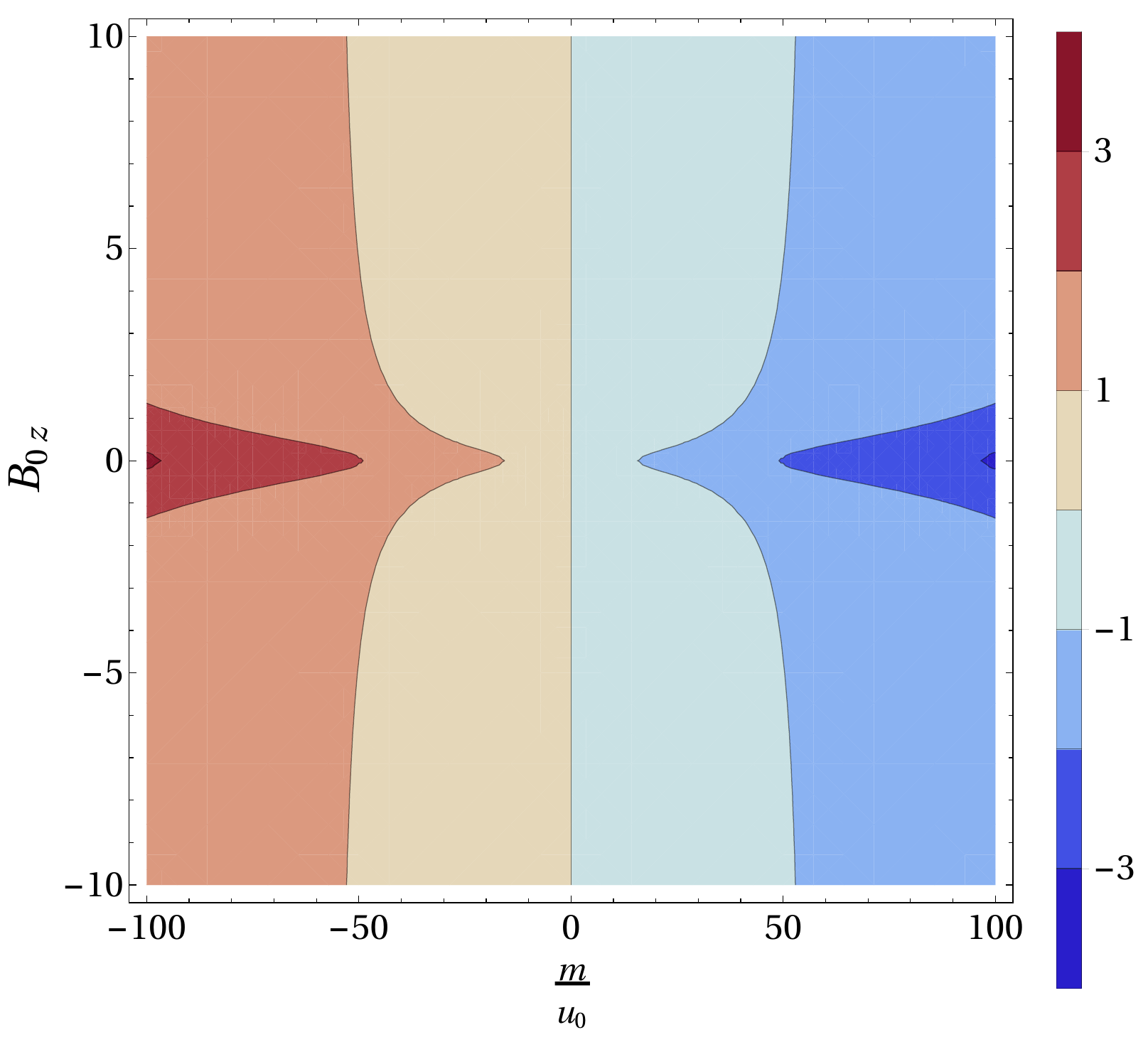}
\caption{Variation of $Im(\beta)$ of the {`fourth solution'} of equation (\ref{eq:disp_with_noise_mag}) {as described in 
\S\ref{sec:With_magnetic_field}} as a function of $m/u_0$ and $B_{0z}$ in the larger range for $q = 1.5$ and $\alpha_3 = 
10$ {for inviscid flow}.}
\label{fig:4th_sol_noise_alpha_10_p_100_B_10}
\end{figure}

The {`fourth solution'} of equation (\ref{eq:disp_with_noise_mag}) is almost complementary solution to the first 
solution of 
the same equation. However, there is some difference, i.e., they are not exactly the complex conjugate as the 
coefficients of equation (\ref{eq:disp_with_noise_mag}) are complex quantities. It will be understood if we go through 
FIG.~\ref{fig:4th_sol_noise_alpha_10_p_5_B_1}, which describes variation of $Im(\beta)$ of the fourth solution of 
equation (\ref{eq:disp_with_noise_mag}) as a function of $m/u_0$ and $B_{0z}$ in the larger domain for $q = 1.5$ and 
$\alpha_3 = 10$. Here we notice that for negative $m/u_0$, the growth rate is positive. For smaller $m/u_0$, as the 
magnitude of the magnetic field increases, {only} up to certain range of the magnetic field, the growth rate 
increases. However, 
for the lager $m/u_0$, the above phenomenon reverses, i.e., as the magnitude of the magnetic field increases, 
the flow becomes less unstable. Now, for the larger range{s} of $m/u_0$ and $B_{0z}$, as described by the 
FIG.~\ref{fig:4th_sol_noise_alpha_10_p_100_B_10} it is the noise that takes over the flow. Consequently, the instability 
in the flow 
arises due to the noise.

\section{The Effect of $Re$}
\label{sec:the_effect_of_Re}
Till now, we have not considered the effect of viscous dissipation. Let us explore the effect of the same on the growth 
rates. Consequently, the equation (\ref{eq:before_dis_rel_hydro_noise}) becomes,
\begin{equation}
 \begin{split}
  \left(i \beta \alpha_3^2 -\frac{\alpha^4}{Re}\right)u(0) +\frac{2 i \alpha_3}{q}\zeta(0)+ \frac{i \alpha_3^3  
B_{0z} B_{x}(0)}{4 \pi }=m_1,&\\
   i \alpha_3\left(1-\frac{2}{q}\right)u(0) +\left(\frac{\alpha^2}{Re} - i\beta\right) \zeta(0)  - 
\frac{i}{4\pi}B_{0z}\alpha_3\zeta_B(0)& \\= m_2,&\\
 - i\beta B_x(0)-i B_{0z}\alpha_3 u(0)= 0,\\
  - i\beta\zeta_B(0)-i B_{0z}\alpha_3\zeta(0) -i\alpha_3 B_x(0)= 0.
  \label{eq:before_dis_rel_hydro_noise_with_re}
 \end{split}
\end{equation}
The corresponding dispersion relation for $m_1 = m_2=m$ from equation (\ref{eq:before_dis_rel_hydro_noise_with_re}) 
becomes,
\begin{equation}
 \begin{split}
  &\frac{m}{u_0} \left(4 \pi\alpha_3^2 B_{0z}^2 \beta -\frac{16 i \pi^2 \alpha_3^2 \beta^2}{Re} 
-16\pi^2\beta^3\right) = \frac{8i\pi B_{0z}^2\alpha_3^4}{q} 
\\&-i\alpha_3^6 B_{0z}^4 +\frac{32\pi^2 \alpha_3 \beta^2}{q}\left(\frac{m}{u_0}\right) +\frac{64i \pi^2 
\alpha_3^2\beta^2}{q^2}\\& -\frac{32i\pi^2\alpha_3^2\beta^2}{q} + 8i\pi\alpha_3^4\beta^2 B_{0z}^2 +\frac{16i\pi^2 
\alpha_3^6 \beta^2}{Re^2} +\frac{32\pi^2 \alpha_3^4 \beta^3}{Re}\\ & -16i\pi^2 \alpha_3^2 
\beta^4.
\label{eq:disp_with_noise_mag_with_re}
 \end{split}
\end{equation}
If the magnetic field is not there in the system, then the above dispersion relation becomes,
\begin{eqnarray}
 \begin{split}
- i\alpha_3^2\beta^2 + \left(\frac{2\alpha_3^4}{Re}+\frac{m}{u_0}\right)\beta+\frac{2 
\alpha_3}{q}\left(\frac{m}{u_0}\right)\\+\frac{4i\alpha_3^2}{q^2} 
-\frac{2i\alpha_3^2}{q} +\frac{i\alpha_3^2}{Re}\left(\frac{m}{u_0}\right) +\frac{i\alpha_3^6}{Re^2} = 0
\label{eq:disp_with_noise_without_mag_with_re}
 \end{split}
\end{eqnarray}
with the solutions
\begin{equation}
 \begin{split}
 &\beta =\\ &\frac{1}{2\alpha_3^2}\left[-i \left(\frac{m}{u_0}\right) -\frac{2i\alpha_3^4}{Re} \pm 
\sqrt{\frac{8\alpha_3^4}{q^2}(2-q)-\frac{m}{u_0}\left(\frac{m}{u_0}+\frac{8i\alpha_3^3}{q}\right)}\right].
\label{eq:hydro_sol_disp_without_mag_with_re}
\end{split}
\end{equation}
If we compare equation (\ref{eq:hydro_sol_disp}) {with equation} (\ref{eq:hydro_sol_disp_without_mag_with_re}), we 
notice 
that the latter has an additional term involving $Re$. Since, the term is imaginary, it reduces the instability in the 
flow. Hence, inclusion of $Re$ makes the flow less unstable if there is no effect of magnetic field in the flow. Equation 
(\ref{eq:disp_with_noise_without_mag_with_re}) is same as equation (9) in {\citealt{Nath_2016}}. Figure 2 in 
{\citealt{Nath_2016}}
also suggests that inclusion of $Re$, decreases the growth rate. 

\begin{figure}		
\includegraphics[width = \columnwidth]{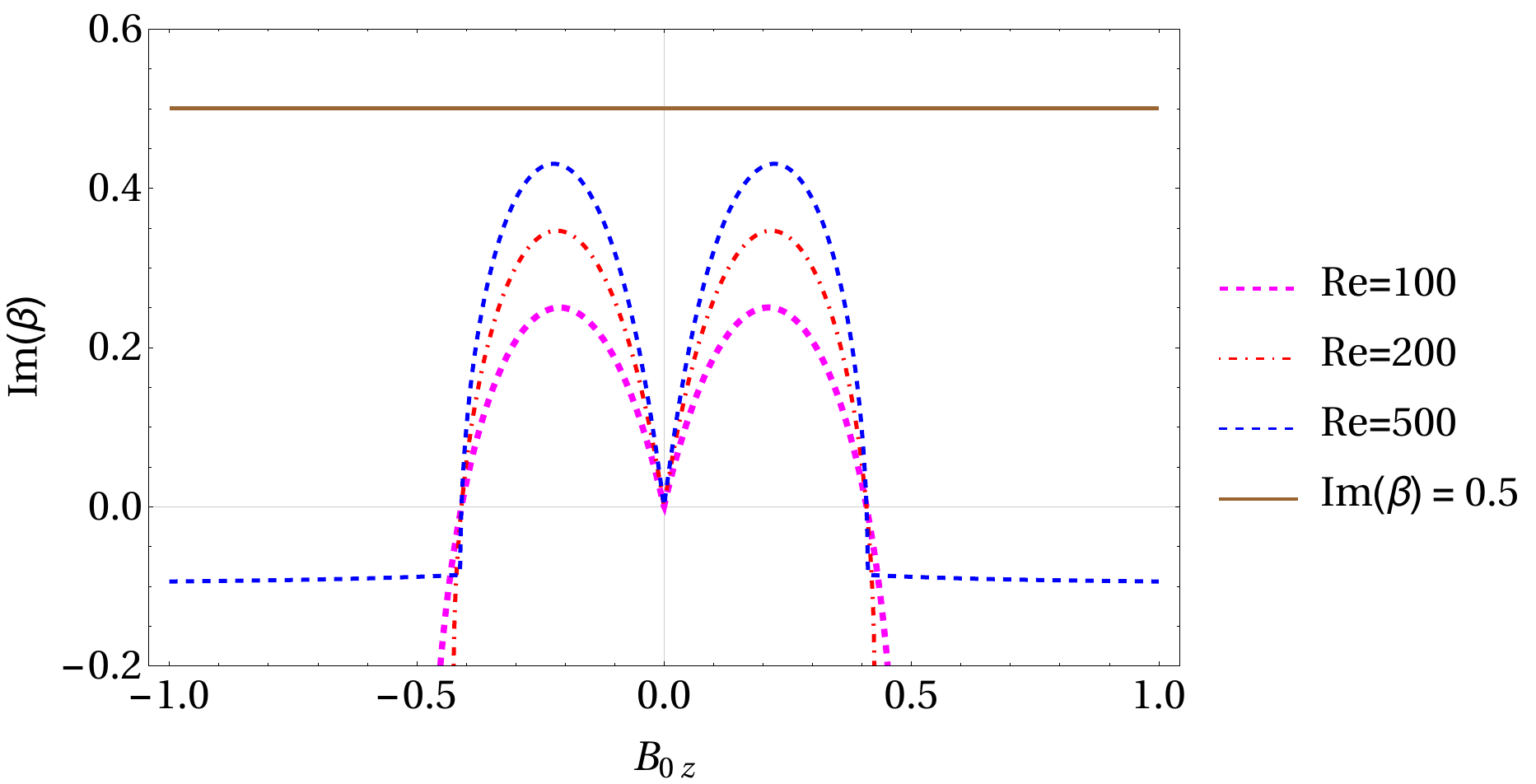}
\caption{Variation of $Im(\beta)$ of the third solution of equation (\ref{eq:disp_with_noise_mag_with_re}) when $m =0$ as 
a function of $B_{0z}$ for $q = 1.5$, $\alpha_3 = 10$ and $Re = 100,\ 200\ \rm{and}\ 500$.}
\label{fig:MRI_diff_alpha_3_diff_re}
\end{figure}

\begin{figure}		
\includegraphics[width = \columnwidth]{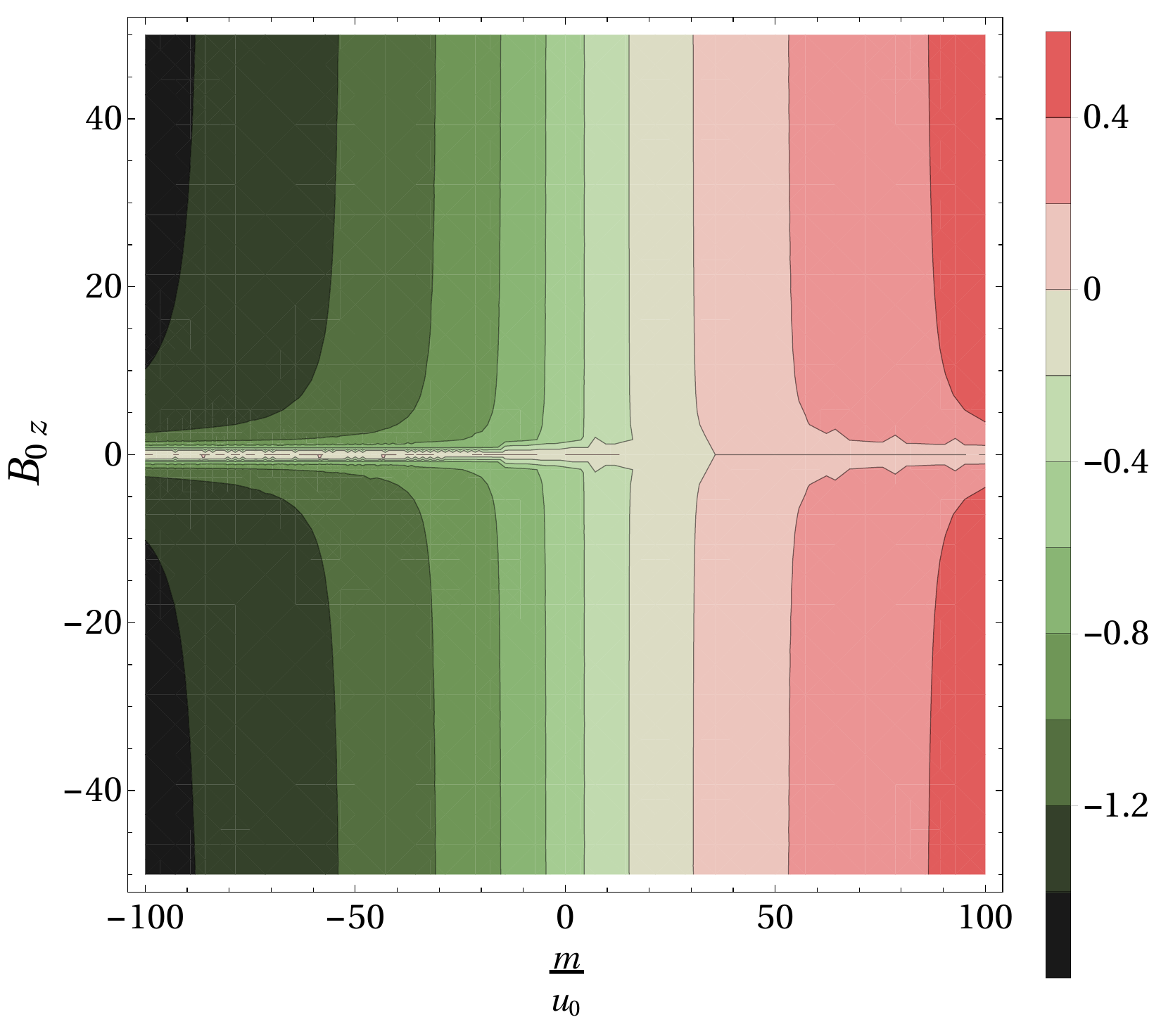}
\caption{Variation of $Im(\beta)$ of the third solution of equation (\ref{eq:disp_with_noise_mag_with_re}) as a function 
of $m/u_0$ and $B_{0z}$ for $q = 1.5$, $\alpha_3 = 10$ and $Re = 100$.}
\label{fig:3rd_sol_noise_alpha_10_p_100_B_50_re_100}
\end{figure}

Let us incorporate the magnetic field in the flow and get rid of the effect of noise. Then we obtain the usual MRI but 
with the affect of viscosity. FIG.~\ref{fig:MRI_diff_alpha_3_diff_re} shows the variation of $Im(\beta)$ of the third 
solution of equation (\ref{eq:disp_with_noise_mag_with_re}) when $m =0$ as a function of $B_{0z}$ for $q = 1.5$, 
$\alpha_3 = 10$ and $Re = 100,\ 200\ \rm{and}\ 500$. This figure shows that {with} the inclusion of viscosity i.e. 
$Re$ in the 
flow, the growth rate decreases. As $Re$ increases the corresponding growth rates also increase and ultimately reach 
{the results of} inviscid limit. This phenomenon becomes obvious once we compare between the 
FIGs.~\ref{fig:MRI_diff_alpha_3_diff_re} and 
\ref{fig:MRI_diff_alpha_3}.  Let us study the effect of $Re$ on the growth rate if both magnetic field and the noise are 
present in the flow. FIG.~\ref{fig:3rd_sol_noise_alpha_10_p_100_B_50_re_100} shows the variation of $Im(\beta)$ of the 
third solution of equation (\ref{eq:disp_with_noise_mag_with_re}) as a function of $m/u_0$ and $B_{0z}$ for $q = 1.5$, 
$\alpha_3 = 10$ and $Re = 100$. If we compare between FIGs.~\ref{fig:3rd_sol_noise_alpha_10_p_100_B_50} and 
\ref{fig:3rd_sol_noise_alpha_10_p_100_B_50_re_100}, we notice that in the later case the growth rate decreases due to 
considering the viscosity, i.e. $Re$ in the flow. 

\begin{figure}		
\includegraphics[width = \columnwidth]{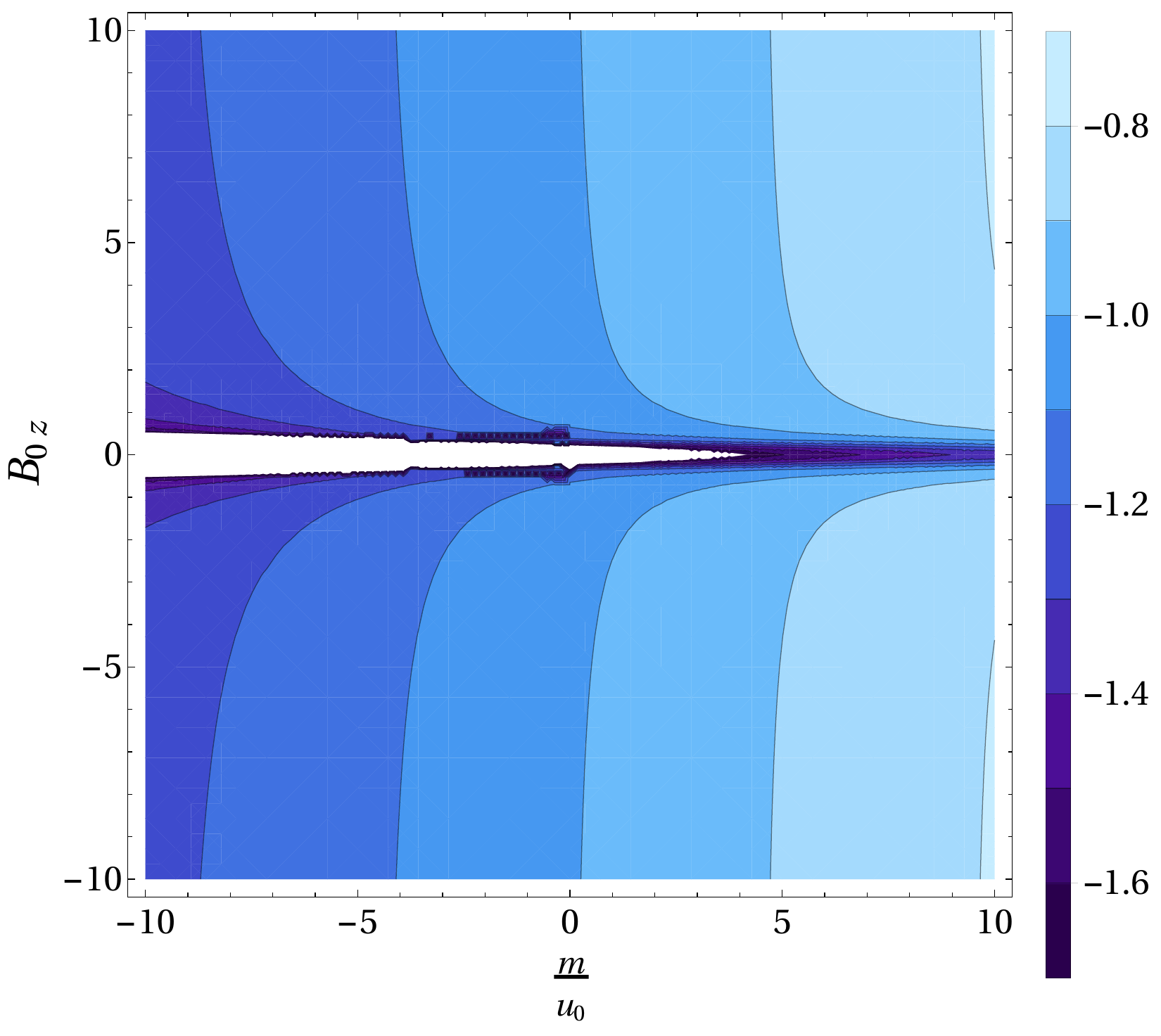}
\caption{Variation of $Im(\beta)$ of the first solution of equation (\ref{eq:disp_with_noise_mag_with_re}) as a function 
of $m/u_0$ and $B_{0z}$ for $q = 1.5$, $\alpha_3 = 10$ and $Re = 50$.}
\label{fig:1st_sol_noise_alpha_10_re_50}
\end{figure}

FIG.~\ref{fig:1st_sol_noise_alpha_10_re_50} shows the variation of $Im(\beta)$ of the first solution of equation 
(\ref{eq:disp_with_noise_mag_with_re}) as a function of $m/u_0$ and $B_{0z}$ for $q = 1.5$, $\alpha_3 = 10$ and $Re = 
50$. If we compare it with FIG.~\ref{fig:1st_sol_noise_alpha_10_p_100_B_10}, we notice that the growth rate becomes 
negative, i.e. the flow becomes stable, for $Re = 50$. However, as $Re$ increases, the growth rates increases and 
eventually at large $Re$, FIG.~\ref{fig:1st_sol_noise_alpha_10_re_50} 
becomes FIG.~\ref{fig:1st_sol_noise_alpha_10_p_100_B_10}. Now, $Im(\beta)$ for the second and fourth solutions of 
equation(\ref{eq:disp_with_noise_mag_with_re}), also get reduced due the presence of the viscosity, i.e. $Re$.

\section{Conclusion}
\label{sec:Conclusion}
The transport of angular momentum {outward} and that of the matter {inward} in the accretion disk is still 
debatable and 
arguably open question till today, particularly at the low-temperature region of the accretion disks. In the hot region, 
MRI is quite successful in explaining the transport. However, there are different parameter regimes (e.g., large magnetic 
field and huge Reynolds number \cite{Nath_2015}, a large toroidal component of the magnetic field \cite{Das_2018}, etc.) 
where MRI gets suppressed or ceases to work. In the previous publications\cite{Ghosh_2020, Ghosh_2021_a, Ghosh_2021_b}, 
we attempted to propose a generic instability mechanism from the hydrodynamical point of view if noise is present in the 
corresponding flow. With that thought, here we {have attempted} to compare the growth {rates} 
corresponding to both the 
mechanisms. We 
{have attempted} to check what happens to the growth rates if both magnetic field as well as the noise with non-zero 
mean is 
present in the flow. From the analysis, our conclusions are following.
\begin{itemize}
	\item Among the four solutions {for the growth rates, one of them reduces to} the MRI growth 
rate 
		if the noise is absent. If both noise as well as the magnetic field are present in the flow, in the MRI 
active region of 
the magnetic field, the growth rates follow the MRI growth rate pattern, i.e., the growth rate first increases, attains a 
maximum, and then again decreases with the variation of magnetic field. However, the corresponding maximum growth rate is 
slightly larger than the MRI growth rate. In the larger domain of magnetic field where MRI is inactive, the instability 
is almost entirely due to the noise as the corresponding growth rates are almost independent of the magnetic field. 
	{However, there is another growth rate which is} almost complementary to the 
	{above mentioned growth rate.} 
\item {Further, there is another solution for the growth rate which reduces} to the growth rate related 
	to the hydrodynamic instability due to the presence of noise in the flow when the magnetic field is absent. 
		This solution has nothing to do with MRI. 
When both magnetic field and noise are retained in the flow, at the MRI active magnetic field region also the growth rate 
is larger than that of the MRI. This growth rate does not have any MRI imprint. The magnitude of the growth rate mostly 
		depends on the mean of the noise. As the mean increases, the growth rate increases. 
		{However, this also has an almost complimentory growth rate, which is the last solution
		in the set of four.}
	\item {Above points summarize that} even in the MRI active region, {as there are two growth
		rates which are larger} at the comparatively larger mean of the noise than those of the other 
		two solutions which have the imprint of MRI, the 
corresponding flow becomes unstable due to the hydrodynamical instability in the presence of noise.
\item All the growth rates decrease due to the presence of viscosity in the flow. As the Reynolds number {increases}, 
the 
corresponding growth rates increase, as expected.
\end{itemize}

\section*{Acknowledgement} S.G. thanks DST India for INSPIRE fellowship. We want to thank Prof. Vinod Krishan 
of the Indian Institute of Astrophysics for her constructive comments and suggestions that help to shape our thoughts 
in better ways. This work is partly supported by a fund of the Department of Science and Technology (DST-SERB) with 
research Grant No. DSTO/PPH/BMP/1946 (EMR/2017/001226) and a fund of the Indian Institute of Science.


%
%

%


\bibliography{magnetic_WKB}{}

\end{document}